\documentclass[12pt]{article}
\usepackage{epsfig}
\usepackage{a4,isolatin1}
\usepackage{amsmath,amsfonts,latexsym, amssymb}

\newtheorem{satz}{Theorem}[section]
\newtheorem{defi}[satz]{Definition}

\newtheorem{bem}[satz]{Remark}
\newtheorem{lemma}[satz]{Lemma}
\newtheorem{koro}[satz]{Corollary}

\newtheorem{conclusion}[satz]{Conclusion}
\newtheorem{ob}[satz]{Observation}

\newtheorem{propo}[satz]{Proposition}

\newcommand{\mcal}{\mathcal}

\newcommand{\tit}{\textit}

\newcommand{\N}{\mathbb{N}}
\newcommand{\R}{\mathbb{R}}
\newcommand{\Z}{\mathbb{Z}}
\newcommand{\bewende}{$ \hfill \Box $}

\begin{document}
\thispagestyle{empty}
\begin{center}
\vspace*{1.0cm}

{\LARGE{\bf A Geometric Renormalisation Group\\ in 
Discrete Quantum Space-Time}}

\vskip 1.5cm

{\large {\bf Manfred Requardt}}\\email: requardt@theorie.physik.uni-goettingen.de 

\vskip 0.5 cm 

Institut f\"ur Theoretische Physik \\ 
Universit\"at G\"ottingen \\ 
Bunsenstrasse 9 \\ 
37073 G\"ottingen \quad Germany

\end{center}

\vspace{1 cm}

\begin{abstract}
  We model quantum space-time on the Planck scale as dynamical
  networks of elementary relations or time dependent random graphs,
  the time dependence being an effect of the underlying dynamical
  network laws. We formulate a kind of geometric renormalisation group
  on these (random) networks leading to a hierarchy of increasingly
  coarse-grained networks of overlapping lumps. We provide arguments
  that this process may generate a fixed limit phase, representing our
  continuous space-time on a mesoscopic or macroscopic scale, provided
  that the underlying discrete geometry is critical in a specific
  sense (geometric long range order).

Our point of view is corroborated by a series of analytic and
numerical results, which allow to keep track of the geometric changes,
taking place on the various scales of the resolution of space-time. Of
particular conceptual importance are the notions of dimension of such
random systems on the various scales and the notion of geometric criticality.

\end{abstract} \newpage
\setcounter{page}{1}

\section{Introduction}
Among the various approaches to \tit{quantum gravity} (or (quantum)
space-time physics) there exists one which assumes that physics and,
in particular, space-time itself are basically discrete on the
presumed fundamental Planck level. This working philosophy is shared
by a variety of more or less related research programs which, however,
employ different technical concepts and follow different lines of
reasoning when it comes to the concrete realisation of such a program
(for a small and incomplete list of papers of other groups see e.g.
\cite{Fot1} to \cite{Anton}, \cite{Is}, \cite{Isham2}, \cite{Raptis}
and \cite{Hooft1} to \cite{Hooft3}, for further references see below).

Our own approach has been developed in \cite{1} to \cite{Quantum}. It
generalizes the concept of \tit{cellular automata} to so-called
\tit{cellular networks} which live on, in general, very large
irregular and dynamical \tit{graphs}. That is, both the nodes
\tit{and} the bonds are assumed to be dynamical degrees of freedom and
interact with each other. An important ingredient of the
\tit{dynamical laws} is the possibility that bonds are switched on and
off in the course of network evolution so that also the overall wiring
or the geometry of the global network is a dynamically changing
structure.

If one starts from such discrete model theories, two important points
are the following. First, the definition of a (class of) primordial
dynamics, which, in one way or the other, have the potential to lead
to our wellknown effective (causal) dynamical evolution laws on an
emergent continuum space-time. Second (and closely related to the
first problem), the control of this continum limit as a limit of a
sequence of increasingly coarse grained intermediate theories. That
is, one of the central issues is it, to reconstruct and recover the
ordinary continuum physics and mathematics, starting from the remote
Planck level. Some steps in this direction have been made in the above
mentioned papers. They depend of course crucially on the kind of model
theory being adopted and the general working philosophy.

In the following we will develop a kind of \tit{geometric
  renormalisation process} leading, as we hope, in the end to a fixed
point (or rather, phase), representing some continuum theory. Our
renormalisation scheme carries the flavor of our particular framework,
that is, the global structure and large scale patterns, existing in
large networks and graphs. In some qualitative sense it is inspired by
the \tit{real-space block variable} approach to renormalisation in the
critical regime of statistical mechanics. One should however note that
the implementation of such a program on the Planck scale is
necessarily much more involved and ambitious as compared to the
typical scales of standard physics. The reason is that both the
patterns, living in the ambient network or space and the background
space itself have to be renormalized, and it turns out to be an
ambitious enterprise to keep track of the relevant geometric changes
and characteristics on the various scales of resolution of space-time.
In particular, among other things, also the dimension of the
underlying spaces will change in
general during the renormalisation process.\\[0.3cm]
Remark: We want to emphasize that, in the absence of a fixed
background space, the clue consists of performing the renormalisation
steps in an \tit{intrinsic} way, without referring to some embedding
space or other external geometric concepts. On the other hand, the
technical methods being developed are expected to be useful also in
other areas of modern physics and can be employed in other
coarse-graining schemes, for example in the field of \tit{dynamical
  triangulation} and \tit{simplicial complexes}.\vspace{0.3cm}

Before we begin with the discussion of the technical details of our
program, we want to add some remarks about the wider physical context
to which such ideas may belong. Illuminating ideas about discreteness
on a fundamental level have already been entertained by Wheeler et al,
see e.g. the last pages in \cite{Wheeler} or
\cite{Wheeler2},\cite{Wheeler3} respectively, another early source is
Myrheim, \cite{Myr}.  Discrete structures like partial orders have for
example been treated by Isham and coworkers, \cite{Is}. A broad and
general approach towards discrete physics in general has been
developed by T.D.Lee and his group (for a collection see vol.3 of his
selected papers, \cite{Lee}). Last but not least, there is the huge
body of work subsumed under the catchword \tit{random geometry} or
\tit{dynamical triangulation} (\cite{Ambjorn} or \cite{Ambjorn2})
which is however mostly concerned with the discretisation of a
preexisting continuous initial manifold.  There may be interesting
connections between our framework and these other approaches but, for
the time being, we refrain from commenting on them in this paper to
keep our paper within reasonable length.

As a last point we want to mention some interesting
cross-fertilisation. In the papers mentioned above we based our
analysis on a class of dynamical network laws which incorporate a
mutual interaction between the local states defined on the nodes of
the underlying graph and the near by bonds. This allows us to treat
both the dynamics of the ordinary degrees of freedom \tit{on} the
graph and the dynamical change of the geometry of the network on the
same footing.

We recently observed that similar ideas have been entertained within
the framework of cellular automata (see e.g. \cite{Ila1} and
\cite{Ila2}), the models being called \tit{structurally dynamic
  cellular automata} or SDCA. As far as we can see at the moment, the
adopted technical framework is not exactly the same but we think, a
comparison of both approaches should turn out to be profitable. We
conclude this introduction with a brief description of what we are
going to do in the following.

In the next section we explain the basics of the framework we are
employing. In section 3 we briefly introduce the concept of a
\tit{random graph}. To establish some contact to other existing
approaches, we show in section 4 that our network naturally carries
also the structure of \tit{causal sets}. The concrete construction of
the renormalisation steps towards an envisaged continuum theory begins
with section 5 which contains also a series of rigorous analytical and
numerical results which are of technical relevance in the subsequent
reasoning. In section 6 we study some simple toy models which (despite
of their simplicity) show that there indeed do exist fixed points in
the category of infinite graphs under our geometric renormalisation
process.  In section 7 we study the behavior of the particularly
important geometric concept of \tit{graph} or \tit{network dimension}
and its behavior under renormalisation and in section 8, which is kind
of a conclusion, we analyse the kind of \tit{geometric criticality}
which is in our view essential in order to arrive at non-trivial
macroscopic limit space-times.

We recently came upon a beautiful discussion of some work of Gromov
(\cite{Berger}), which shows that there may be some deep and
interesting connections between our framework, developed in the
following, and ideas of coarse graining in, for example, geometric
group theory by Gromov (see also the references
\cite{Harpe},\cite{Bartholdi},\cite{Grigorchuk} cited in section 7).

\section{Protogeometry and Protodynamics}

In a first step we want to motivate why we choose exactly the kind of
model theory, we are discussing in the following. On the one side we
have a working philosophy which is similar to the one, expounded by 't
Hooft in e.g. \cite{Hooft1} to \cite{Hooft3}. That is, we entertain
the idea that for example quantum theory may well emerge as an
\tit{effective (continuum) theory} on the mesoscopic scale of an
underlying discrete more microscopic theory. As we want our underlying
(pre)geometry to coevolve with the patterns living in this substratum,
we developed the above mentioned generalisation of the more regular
cellular automata.

Another essential property of such \tit{discrete dynamical systems} is
, while the basic ingredients and elementary building blocks are
reasonably simple, their potential for the emergence of very complex
behavior on the more macroscopic scales, thus supporting the speculation
that such systems may be capable of generating viable continuum
theories.

We now begin to introduce the necessary technical ingredients. We
start with the definition of some notions of graph theory.  
\begin{defi}\label{graph1}A simple, countable, labelled, undirected graph, $G$, consists of a countable set of nodes or vertices, $V$, and a set of
  edges or bonds, $E$, each connecting two of the nodes. There exist no
  multiple edges (i.e. edges, connecting the same pair of nodes) or
  elementary loops (a bond, starting and ending at the same node). In
  this situation the bonds can be described by giving the
  corresponding set of unordered pairs of nodes. The members of $V$
  are denoted by $x_i$, the bonds by $e_{ij}$, connecting the nodes
  $x_i$ and $x_j$.
\end{defi}
Remarks: We could also admit a non-countable vertex set. The above
restriction is only made for technical convenience. From a physical
point of view one may argue that the \tit{continuum} or uncountable
sets are idealisations, anyhow. The notions \tit{vertex, node} or
\tit{edge, bond} are used synonymously. Furthermore, the labeling of
the nodes is only made for technical convenience (to make some
discussions easier) and does not carry a physical meaning. As in
general relativity, all models being invariant under \tit{graph
  isomorphisms} (i.e. relabelling of the nodes and corresponding
bonds) are considered to be physically equivalent.
\vspace{0.3cm}

In the above definition the bonds are not directed (but oriented; see
below). In certain cases it is also useful to deal with directed
graphs.
\begin{defi}A directed graph is a graph as above, with $E$ consisting
  now of directed bonds or ordered pairs of nodes. In this case we
  denote the edge, pointing from $x_i$ to $x_j$ by $d_{ij}$. There
  may also exist the opposite edge, denoted by $d_{ji}$.
\end{defi}
\begin{ob}An undirected graph, as in definition \ref{graph1}, can be
  considered as a particular directed graph with $e_{ij}$
  corresponding to the pair of directed edges, $d_{ij},d_{ji}$.
\end{ob}
\begin{bem}We introduced and studied algebraic and functional analytic
  structures like e.g. Hilbert spaces and Dirac operators on such
  graphs in \cite{3},\cite{5}. In such situations, the bonds,
  $e_{ij}$, $d_{ij}$, can be given a concrete algebraic meaning with
\begin{equation}e_{ij}:=d_{ij}-d_{ji}=-e_{ji}\end{equation}
\end{bem} 

It is now suggestive to regard the edges between pairs of points as
describing their (direct) interaction. This becomes more apparent if
we impose dynamical network laws on these graph structures so that
they become a particular class of discrete dynamical systems.
Henceforth we denote such a dynamical network, which is supposed to
underly our continuous space-time manifold, by $QX$ (``\tit{quantum
  space}''). We want to make the general remark that the \tit{cellular
  networks}, introduced in the following, can either be regarded as
mere models of a perhaps more hypothetical character, encoding, or
rather simulating, some of the expected features of a surmised
\tit{quantum space-time} or, on the other hand, as a faithful
realisation of the primordial substratum, underlying our macroscopic
space-time picture.  Up to now, this is a matter of
taste.\vspace{0.3cm}

For technical convenience and to keep matters reasonable simple, we
choose a discrete overall clock-time (not to be confused with the
\tit{physical time} which is rather supposed to be an emergent and
intrinsic characteristic, related to the evolution of
quasi-macroscopic patterns in such large and intricately wired
networks). In principle the clock-time can also be made into a local
dynamical variable. Furthermore, we assume the node set of our initial
network to be fixed and being independent of clock-time (in contrast
to the bonds). We will see in the following sections that this feature
will change under the renormalisation steps, i.e. on the highler
levels, the class of lumps may become dependent on time.

We assume that each node, $x_i$, or bond, $e_{ik}$, carries an
internal (for simplicity) discrete \tit{state space}, the internal
states being denoted by $s_i$ or $J_{ik}$. In simple examples we chose
for instance:
\begin{equation} s_i \in q\cdot \mathbb{Z}\quad,\quad J_{ik}\in
  \{-1,0,+1\} \end{equation}
with $q$ an elementary quantum of information and
\begin{equation}e_{ki}=-e_{ik}\Rightarrow J_{ki}=-J_{ik}  \end{equation}
In most of the studied cellular automata systems even simpler internal
state spaces are chosen like e.g. $s_i\in \{0,1\}$. This is at the
moment not considered to be a crucial point. The above choice is only
an example.

In our approach the bond states are dynamical degrees of freedom
which, a fortiori, can be switched off or on (see below). Therefore
the \tit{wiring}, that is, the pure \tit{geometry} (of relations) of
the network is a clock-time dependent, dynamical property and is
\tit{not} given in advance.  Consequently, the nodes and bonds are
typically not arranged in a more or less regular array, a regular lattice say,
with a fixed nea-/far-order.  This implies that \tit{geometry} will
become to some degree a \tit{relational} (Machian) concept and is no
longer a static background.

As in cellular automata, the node and bond states are updated (for
convenience) in discrete clock-time steps, $t=z\cdot\tau$,
$z\in\mathbb{Z}$ and $\tau$ being an elementary clock-time interval.
This updating is given by some \tit{local} dynamical law (examples are
given below). In this context \tit{local} means that the node/bond
states are changed at each clock time step according to a prescription
with input the overall state of a certain neighborhood (in some
topology) of the node/bond under discussion.

A simple example of such a local dynamical law we are having in mind
is given in the following definition (first introduced in \cite{3}).
\begin{defi}[Example of a Local Law]
At each clock time step a certain {\em quantum} $q$
is exchanged between, say, the nodes $x_i$, $x_k$, connected by the
bond $e_{ik}$ such that 
\begin{equation} s_i(t+\tau)-s_i(t)=q\cdot\sum_k
  J_{ki}(t)\end{equation}
(i.e. if $J_{ki}=+1 $ a quantum $q$ flows from $x_k$ to $x_i$ etc.)\\
The second part of the law describes the {\em back reaction} on the bonds
(and is, typically, more subtle). We assume the
existence of two {\em critical parameters}
$0\leq\lambda_1\leq\lambda_2$ with:
\begin{equation} J_{ik}(t+\tau)=0\quad\mbox{if}\quad
  |s_i(t)-s_k(t)|=:|s_{ik}(t)|>\lambda_2\end{equation}
\begin{equation} J_{ik}(t+\tau)=\pm1\quad\mbox{if}\quad 0<\pm
  s_{ik}(t)<\lambda_1\end{equation}
with the special proviso that
\begin{equation} J_{ik}(t+\tau)=J_{ik}(t)\quad\mbox{if}\quad s_{ik}(t)=0
\end{equation}
On the other side
\begin{equation} J_{ik}(t+\tau)= \left\{\begin{array}{ll} 
\pm1 & \quad J_{ik}(t)\neq 0 \\
0    & \quad J_{ik}(t)=0
\end{array} \right. \quad\mbox{if}\quad
\lambda_1\leq\pm
  s_{ik}(t)\leq\lambda_2 
\end{equation}
In other words, bonds are switched off if local spatial charge
fluctuations are too large or switched on again if they are too
small, their orientation following the sign of local charge
differences, or remain inactive.

Another interesting law arises if one exchanges the role of
$\lambda_1$ and $\lambda_2$ in the above law, that is, bonds are
switched off if the local node fluctuations are too small and are
switched on again if they exceed $\lambda_2$.
\end{defi}
We make the following observation:
\begin{ob}[Gauge Invariance] The above dynamical law depends nowhere on the 
  absolute values of the node ``charges'' but only on their relative
  differences. By the same token, charge is nowhere created or
  destroyed. We have
\begin{equation}\Delta(\sum_{QX}s(x))=0\end{equation}
($\Delta$ denoting the change in total charge of the network between
two consecutive clocktime steps).
To avoid artificial ambiguities we can e.g. choose a fixed reference
level, taking as initial
condition at $t=0$ the following constraint 
\begin{equation}\sum_{QX}s(x)=0\end{equation} 
\end{ob}

We resume what we consider to be the crucial ingredients of network
laws, we are interested in
\begin{enumerate}
\item As in gauge theory or general relativity, our evolution law
  should implement the mutual interaction of two fundamental
  substructures, put a little bit vaguely : ``{\em geometry}'' acting
  on ``{\em matter}'' and vice versa, where in our context ``{\em
    geometry}'' is assumed to correspond in a loose sense to the local
  and/or global array of bond states and ``{\em matter}'' to the
  structure of the node states.
\item By the same token the alluded {\em selfreferential} dynamical
  circuitry of mutual interactions is expected to favor a kind of {\em
    undulating behavior} or {\em selfexcitation} above a return to
  some uninteresting {\em equilibrium state} (being devoid of stable
  structural details), as is frequently the case in systems consisting
  of a single component which directly acts back on itself. This
  propensity for the {\em autonomous} generation of undulation
  patterns is in our view an essential prerequisite for some form of
  ``{\em protoquantum behavior}'' we hope to recover on some coarse
  grained and less primordial level of the network dynamics.
\item In the same sense we expect the large scale pattern of switching-on and
 -off of bonds to generate a kind of ``{\em protogravity}''.
\end{enumerate}
Remark: The above dynamical law shows that bonds with $J_{ik}=0$ at
clock time $t$ do not participate in the dynamics in the next time
step. We hence may consider them as being temporally inactive. The
shape of the network, neglecting all the internal states on the nodes
and bond together with the inactive bonds we call the \tit{wiring
  diagram}.\vspace{0.3cm}

If one concentrates solely on this \tit{ wiring diagram}, figure 1
describes one clocktime step in the life of a \tit{dynamic graph}. In the
picture only a small subgraph is shown and the deletion and creation
of edges (that is, elementary interactions among nodes or possible
information channels). The new bonds are represented as bold lines. It
should be emphasized that the graph is \tit{not} assumed to be a
triangulation of some preexisting smooth manifold. This is emphasized
by the existence of edges, connecting nodes which are not necessarily
close with respect to e.g.  the euclidean distance.
\begin{figure}[h]
\centerline{\epsfig{file=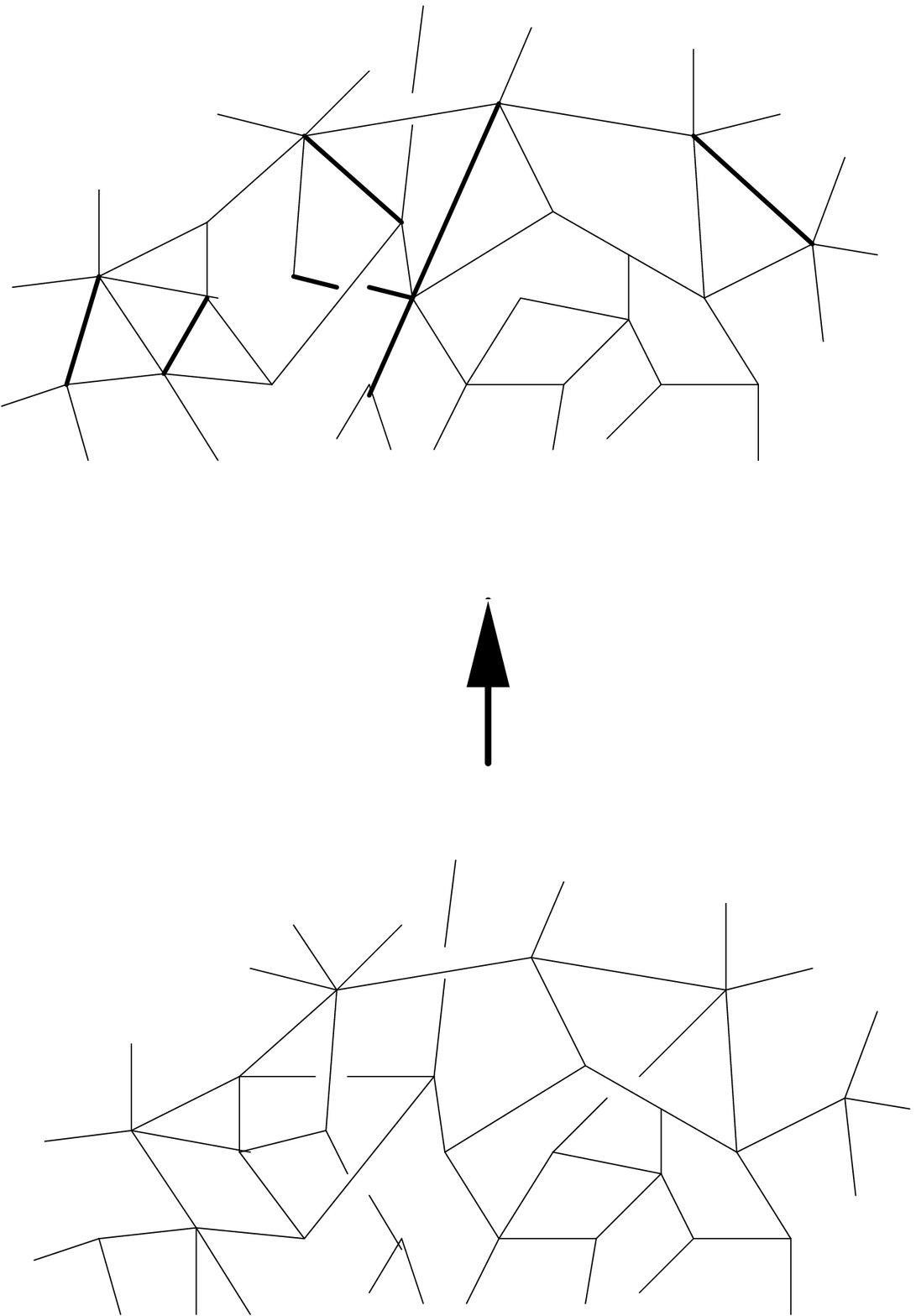,width=7cm,height=7cm,angle=0}}
\caption{}
\end{figure} 

We have pictured our proto space-time on the Planck scale as a
fluctuating network of dynamic relations or exchange of pieces of
information between a given set of nodes. At each fixed clock-time
step there exist in this network certain subclusters of nodes which
are particularly densely entangled and the whole graph can be covered
by this uniquely given set of subclusters of nodes and the respective
induced subgraphs. We dealt with these distinguished clusters of nodes
(called cliques or lumps) in quite some detail in e.g. \cite{4} or
\cite{6}. We emphasize the interesting relations to earlier ideas of
Menger, Rosen et al, which have been discussed in \cite{6}.

One of our core ideas is it that the seemingly structureless
(mathematical) points, making up our ordinary continuous manifolds,
would display a rich nested internal structure if looked at under a
magnification or resolution so that the lumpy structure of space-time
became visible. We think this hidden substructure will become
particularly relevant when it comes to the interpretation of
\tit{quantum phenomena} ( \cite{Quantum}, where possible relations to
some interesting ideas of Connes have been set up).

From a more technical or practical point of view we need a general
principle which allows us to lump together subsets of nodes, living on
a certain level of resolution of space-time, to get the building
blocks of the next level of coarse graining (see below). After a
series of such coarse graining steps we will wind up with a nested
structure of lumps, containing smaller lumps and so forth, which,
after appropriate \tit{rescaling}, may yield in the end some
quasi-continuous but nested structure. This principle is provided by
the following mathematical concept.
\begin{defi}[Subsimplices and Cliques]With $G$ a given fixed graph and
  $V_i$ a subset of its vertex set $V$, the corresponding {\em induced
    subgraph} over $V_i$ (that is, its edges being the corresponding
  edges, occurring in $G$) is called a subsimplex, if all its pairs of
  nodes are connected by a bond. In this class, which is in fact
  partially ordered, the order being given by graph inclusion, there
  exist certain {\em maximal subsimplices}, that is, subsimplices so
  that every addition of another node of the underlying graph(together
  with the respective bonds existing in $G$ (pointing to other nodes
  of the chosen subset) destroys this property.  These maximal
  simplices are usually called {\em cliques} in combinatorics (we call
  them also lumps) and are the candidates for our construction of {\em
    physical points}.  Henceforth we denote them by $C_i$.
  \end{defi}
  
  It has been described in detail in e.g. section 4 of \cite{4} how
  these cliques can be constructed in an algorithmic way, starting
  from an arbitrary node. Note in particular that a given node will,
  in general belong to many different (overlapping) cliques or lumps.
  The situation is illustrated by the following picture:
\begin{figure}[h]
\centerline{\epsfig{file=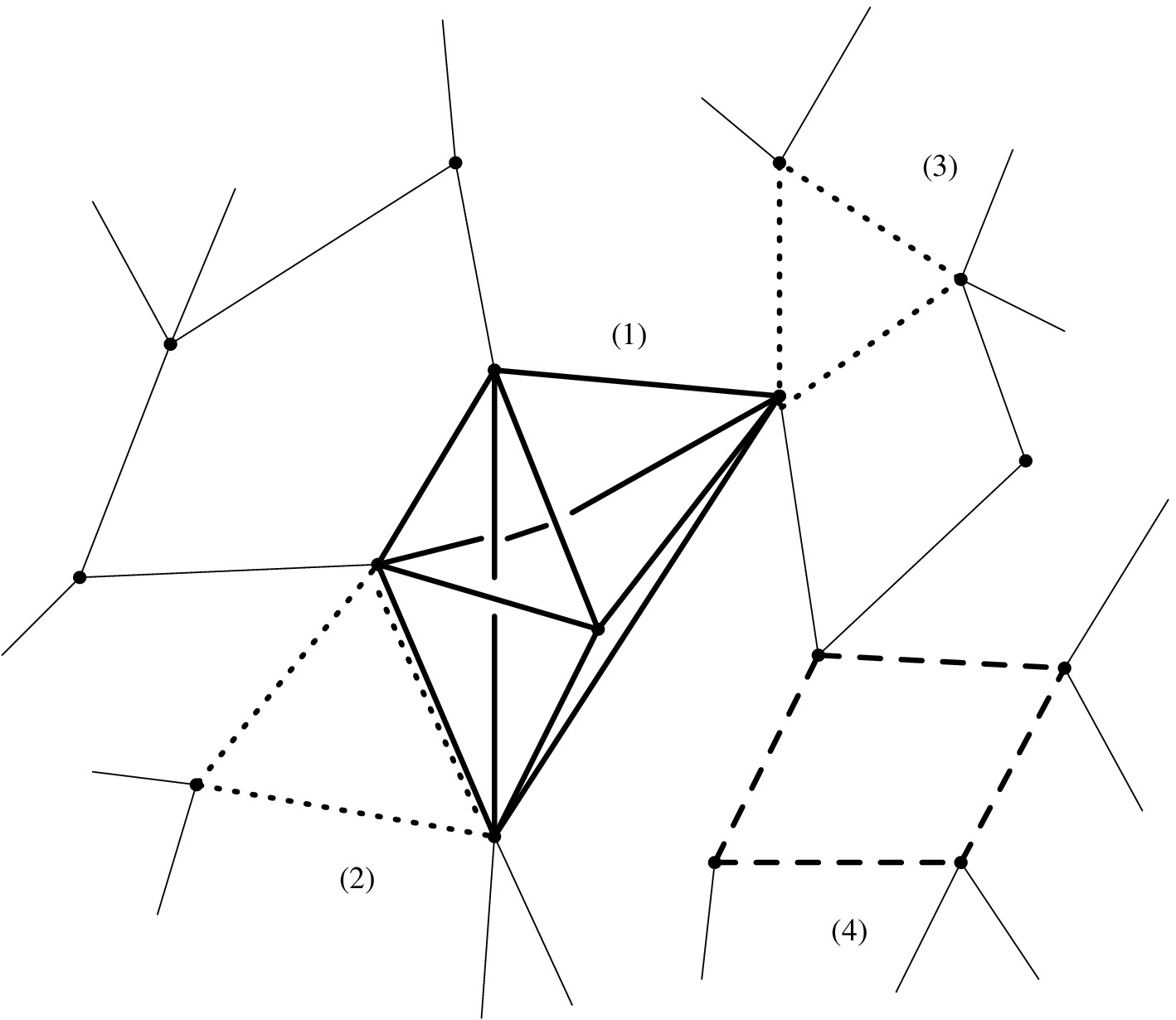,width=6cm,height=6cm,angle=0}}
\caption{}
\end{figure}
In this picture we have drawn a subgraph of a larger graph. $(1)$
denotes a clique, i.e. a maximal subsimplex. Subsets of nodes of such
a clique support subsimplices (called faces in algebraic topology),
the clique being the maximal element in this partial ordered set.
$(2)$ and $(3)$ are other, smaller cliques which overlap with $(1)$ in
a common bond or node. $(4)$ is an example of a subgraph which is not
a clique or subsimplex. Evidently, each node or bond lies in at least
one clique. The smallest possible cliques which can occur in a
connected graph consist of two nodes and the corresponding edge.
\section{\label{cliques} Dynamical Networks as Random Graphs}
\subsection{The Statistical Hypothesis}
As we are dealing with very large graphs, which are, a
fortiori, constantly changing their shape, that is, their distribution
of (active) bonds, we expect the dynamics to be sufficiently stochastic so that
a point of view may be appropriate, which reminds of the working
philosophy of \tit{statistical mechanics}. This does however not
imply that our evolving network is nothing but a simple \tit{random
  graph} as introduced below (cf. the remarks at the end of this section).
It rather means that some of its geometric characteristics can, or
should, be studied within this well-developed context.

Visualizing the characteristics and patterns being prevalent in large
and ``typical'' graphs was already a notorious problem in
\tit{combinatorial graph theory} and led to the invention of the
\tit{random graph} framework (see the more complete discussion in
\cite{4}). The guiding idea is to deal with graphs of a certain
type in a probabilistic sense. This turns out to be particularly
fruitful as many graph characteristics (or their absence) tend to
occur with almost certainty in a probabilistic sense (as has been
first observed by Erd\"os and R\'enyi). The standard source is
\cite{Bollo1} (for further references see \cite{4}).

Another strand of ideas stems from the theory of dynamical systems and
cellular automata, where corresponding statistical and ensemble
concepts are regularly employed. Typically, we are looking for
\tit{attractors} in phase space, which are assumed to correspond to
large scale, that is, after \tit{coarse graining} and \tit{rescaling},
quasi continuous or macroscopic patterns of the system. Experience
shows, that such a structure or the approach towards attractors is in
many cases relatively robust to the choice of initial configurations
or microscopic details and, hence, suggests an ensemble picture.

Furthermore, since the early days of statistical mechanics, the
ensemble point of view (see for example \cite{Gibbs}) is, at least
partly, corroborated by the philosophy that time averages transform
(under favorable conditions) into ensemble averages. In our context
this means the following. Denoting the typical length/time scale of
ordinary quantum theory by $[l_{qm}],[t_{qm}]$, we have
\begin{equation}[l_{qm}]\gg[l_{pl}]\quad,\quad
  [t_{qm}]\gg[t_{pl}]\end{equation}
the latter symbols denoting the Planck scale. Under renormalisation
the mesoscopic scales comprise a huge number of microscopic clock time
intervals and degrees of fredom of the network under discussion.

A fortiori, the networks, we are interested in, correspond to graphs,
having a huge \tit{vertex degree}, i.e. channels, entering a given
typical node of the graph. That is, we expect large local fluctuations
in microscopic grains of space or time. Put differently, the network
locally traverses a large number of different microscopic states in a
typical mesoscopic time interval, $[t_{qm}]$. This observation
suggests that, on a mesoscopic or macroscopic scale, microscopic
patterns will be washed out or averaged over.
\subsection{The Random Graph Framework}
One kind of probability space is constructed as follows. Take all
possible labeled graphs over $n$ nodes as probability space $\cal{G}$
(i.e. each graph represents an elementary event). The maximal possible
number of bonds is $N:=\binom{n}{2}$, which corresponds to the unique
{\em simplex graph} (denoted usually by $K_n$). Give each bond the
{\em independent probability} $0\leq p\leq 1$, (more precisely, $p$ is
the probability that there is a bond between the two nodes under
discussion). Let $G_m$ be a graph over the above vertex set, $V$,
having $m$ bonds. Its probability is then
\begin{equation}pr(G_m)=p^m\cdot q^{N-m}\end{equation}
where $q:=1-p$. There exist $\binom{N}{m}$ different labeled
graphs $G_m$, having $m$ bonds, and the above probability is correctly normalized,
i.e.
\begin{equation}pr({\cal G})=\sum_{m=0}^N\binom{N}{m}p^mq^{N-m}=(p+q)^N=1\end{equation}
This probability space is sometimes called the space of {\em
  binomially random graphs} and denoted by ${\cal G}(n,p)$. Note that
the number of edges is binomially distributed, i.e.
\begin{equation}pr(m)=\binom{N}{m}p^mq^{N-m}\end{equation}
and
\begin{equation}\langle m\rangle=\sum m\cdot pr(m)=N\cdot p\end{equation}

The really fundamental observation made already by Erd\"os and R\'enyi (a
rigorous proof of this deep result can e.g. be found in \cite{Bollo2})
is that there are what physicists would call \tit{phase transitions}
in these \tit{random graphs}. To go a little bit more into the details
we have to introduce some more graph concepts.
\begin{defi}[Graph Properties]{\em Graph properties} are certain
  particular {\em random
    variables} (indicator functions of so-called events) on the above
  probability space ${\cal G}$. I.e., a graph property, $Q$, is
  represented by the subset of graphs of the sample space having the
  property under discussion.
\end{defi}
To give some examples: i) connectedness of the graph, ii) existence
and number of certain particular subgraphs (such as subsimplices
etc.), iii) other geometric or topological graph properties etc.

In this context Erd\"os and R\'enyi made the following important
observation.
\begin{ob}[Threshold Function]A large class of {\em graph properties}
  (e.g. the {\em monotone increasing ones}, cf. \cite{Bollo1} or
  \cite{Bollo2}) have a so-called {\em threshold function}, $m^*(n)$,
  with $m^*(n):=N\cdot p^*(n)$, so that for $n\to\infty$ the graphs
  under discussion have {\em property} $Q$ {\em almost shurely} for
  $m(n)>m^*(n)$ and {\em almost shurely not} for $m(n)<m^*(n)$ or vice
  versa (more precisely: for $m(n)/m^*(n)\to \infty\;\text{or}\;0$;
  for the details see the above cited literature). That is, by turning
  on the probability $p$, one can drive the graph one is interested in
  beyond the phase transition threshold belonging to the graph
  property under study. Note that, by definition, threshold functions
  are only unique up to ``factorization'', i.e. $m^*_2(n)=
  O(m^*_1(n))$ is also a threshold function.
\end{ob}

Calculating these graph properties is both a fascinating and quite
intricate enterprise. In \cite{4} we mainly concentrated on
properties of \tit{cliques}, their distribution (with respect to their
order, $r$, i.e. number of vertices), frequency of occurence of cliques
of order $r$, degree of mutual overlap etc. We then related these
properties to the various assumed stages and phases of our space-time
manifold.

We can introduce various \tit{random function} on the above
probability space.  For each subset $V_i\subset V$ of order $r$ we
define the following random variable:
\begin{equation}X_i(G):=
\begin{cases}1 & \text{if $G_i$ is an $r$-simplex},\\  
 0 & \text{else}
\end{cases}
\end{equation}
where $G_i$ is the corresponding induced subgraph over $V_i$ in $G\in
{\cal G}$ (the probability space). Another random variable is then the
\tit{number of $r$-simplices in $G$}, denoted by $Y_r(G)$ and we have:
\begin{equation}Y_r=\sum_{i=1}^{\binom{n}{r}}X_i\end{equation}
with $\binom{n}{r}$ the number of $r$-subsets $V_i\subset V$. With respect
to the probability measure introduced above we have for the
\tit{expectation values}:
\begin{equation}\langle Y_r \rangle = \sum_i \langle X_i \rangle\end{equation}
and
\begin{equation}\langle X_i \rangle = \sum_{G\in{\cal G}} X_i(G)\cdot
  pr(G_i=\text{$r$-simplex in}\;G).\end{equation} 
These  expectation values were calculated in \cite{4}. We have for example
\begin{equation}\langle X_i \rangle = p^{\binom{r}{2}}\end{equation}

The probability that such a subsimplex is maximal, i.e. is a cliques
is then
\begin{equation}pr(G_r\;\text{is a clique})=(1-p^r)^{n-r}\cdot
  p^{\binom{r}{2}}\end{equation} 
As there exist exactly $\binom{n}{r}$ possible different $r$-sets in
the node set $V$, we arrive at the following conclusion: 
\begin{conclusion}[Distribution of Subsimplices and Cliques]The
  expectation value of the random variable `{\em number of $r$-subsimplices}' is
\begin{equation}\langle Y_r \rangle = \binom{n}{r}\cdot
  p^{\binom{r}{2}}\end{equation}
 For $Z_r$, the {\em number of $r$-cliques} (i.e. maximal! $r$-simplices) in the random graph, we have
 then the following relation
\begin{equation}\langle Z_r
  \rangle=\binom{n}{r}\cdot(1-p^r)^{n-r}\cdot
  p^{\binom{r}{2}}\end{equation}
\end{conclusion}

These quantities, as functions of $r$ (the \tit{order} of the
subsimplices) have quite a peculiar numerical behavior. We are
interested in the typical \tit{order of cliques} occurring in a
generic random graph (where typical is understood in a probabilistic
sense.
\begin{defi}[Clique Number]The maximal order of occurring cliques contained in
  $G$ is called its {\em clique number}, $cl(G)$. It is another random
  variable on the probability space ${\cal G}(n,p)$.
\end{defi}
It is remarkable that this value is very sharply defined in a
typical random graph. Using the above formula for $ \langle
Z_r\rangle$, we can give an approximative value, $r_0$, for this
expectation value and get
\begin{equation}r_0\approx 2\log(n)/\log(p^{-1})+ O(\log\log(n))\end{equation} (cf. chapt. XI.1 of \cite{Bollo1}). It holds that practically all the
occurring cliques fall in the interval $(r_0/2,r_0)$. We illustrate
this with the following tables. Our choice for $n$, the number of
vertices, is $10^{100}$. The reason for this seemingly very large
number is, that we want to deal with systems ultimately simulating our
whole universe or continuous space-time manifolds (see the more
detailed discussion in \cite{4}). We first calculate $r_0$.
\begin{equation}\label{tabelle1}
\begin{array}{|l|c|c|c|c|c|c|c|c|c|}
p & 0.9 & 0.8 & 0.7 & 0.6 & 0.5 & 0.4 & 0.3 & 0.2 & 0.1\\ \hline
r_0 & 4370 & 2063 & 1291 & 901 & 664 & 502 & 382 & 286 & 200
\end{array}
\end{equation}
(for reasons we do not understand we made some numerical errors in the
orriginal table 1 in \cite{4}, p.2043).
 
It is more complicated to give numerical estimates of the distribution
of cliques, that is $\langle Z_r\rangle$. After some manipulations and
approximations we arrived in (\cite{4},p.2051f) at the following
approximative formula and numerical table (the numerical values are
given for $p=0.7$; note that for this parameters the maximal order of
occurring cliques, $r_0$, was approximately $1291$)
\begin{equation}\log(\langle Z_r\rangle)\approx
  r\cdot\log(n)+n\cdot\log(1-p^r)+r^2/2\cdot\log(p)\end{equation}
(with $r^2/2$ an approximation of $r(r-1)/2)$ for $r$ sufficiently large).\\[0.3cm]
\begin{equation} \label{tabelle2}
\begin{array}{|c|c|c|c|c|c|c|c|}
r & 600 & 650 & 800 & 1000 & 1200 & 1300 & 1400\\
\hline
\log(\langle Z_r\rangle) & -5.7\cdot 10^6 & 3.2\cdot
10^4 & 3.2\cdot 10^4 & 2.5\cdot 10^4 & 8.4\cdot 10^3 & -0.75\cdot 10^2 & -1.1\cdot 10^4
\end{array}
\end{equation}
(In the original table 2 of \cite{4} the numerical values for small
and large $r$'s, lying outside the interval $(r_0/2,r_0)$, were wrong as
we neglected numerical contributions which are only vanishingly small
in the above interval. The above table nicely illustrates how fast the
frequency of cliques of order $r$ drops to zero outside the above
interval.

As to the interpretation of these findings, one should remind the
reader that the above results apply to the generic situation, that is,
do hold for typical graphs (in very much the same sense as in
corresponding discussions in the foundations of statistical
mechanics). An evaluation of the combinatorial expressions in this and
the following sections show that frequently the same kind of extreme
probabilistic concentration around, for example, \tit{most probable
values} occurs as in ordinary statistical mechanics. 

What is not entirely clear is, how far the random graph approach can
be applied to our complex dynamical networks. Our working philosophy
is that these results serve to show, what we hope, is the qualitative
behavior of such systems. As our systems follow deterministic
dynamical laws, starting from certain initial conditions, the behavior
cannot be entirely random in the strict sense. This holds the more so
since we expect the systems to evolve towards \tit{attracting sets} in
phase space and/or generate some large scale patterns. On the other
hand, due to the constant reorientation of the bonds, being incident
with an arbitrary but fixed node and the generically large vertex
degrees of the nodes, one may assume that the system is sufficiently
random on small scales, so that the random graph picture reproduces at
least the qualitative behavior of such extremely complex systems.

To make this picture more quantitative, the general strategy is the
following. We count the typical number of active bonds
in our evolving network at a given clock time $t$, calculate from this the
corresponding bond probability, $p(t)$, and relate this snapshot of
our network to a random graph with the same! bond probability. This
should yield at least some qualitative clues. That is, we expect that
qualitative characteristics of our evolving network can, at each given
clock time, be related to the characteristics of a corresponding
random graph. In this specific sense, one may regard the \tit{bond
probability}, $p(t)$, as the crucial \tit{order parameter} of our
network, regarded as a statistical system. (We note that we
implemented such networks on a computer and made detailed studies of
their dynamical behavior and stochastic properties, see
\cite{Nowotny}. Our investigations showed that at least qualitatively
the expected phenomena came out correctly).
\section{Cellular Networks as Causal Sets}
In this section we want to make contact with an approach to quantum
gravity, being based on the concept of \tit{causal sets}. We again
emphasize that, for reasons of technical simplicity, we treat
\tit{time} as a global non-dynamical quantity, being well aware that
this may be a severe restriction. On the other hand, the notorious
so-called \tit{problem of time} has not yet been settled anyhow in
quantum gravity and needs an extra and careful treatment. Under this
proviso we want to show, that our \tit{cellular networks} and
\tit{lump-spaces} automatically have the structure of causal sets,
with this extra structure being induced by our local dynamical laws.
On the most elementary level we start from our above initial network.

We argued above that we want to neglect the details of the (time
dependent) internal states of nodes and bonds and keep only track of
the bonds which are in operation at a given clock-time, $t$, that is,
the bonds with $J_{ik}\neq 0$. Doing this, we arrive at the concept of
\tit{time dependent graphs}, $G(t)$.
\begin{defi}$G(t)$ is a graph with a fixed (time independent) node
  set, $V$, but a time dependent set of active bonds, $E(t)$. In
  principle we could also make the node set time dependent, the above
  assumption is mainly made for convenience.
\end{defi}

The local dynamical laws can as well be viewed as a prescription, by
which local pieces (quanta) of information are transported between the
active bonds of the network. The nodes, which can be reached from a
given node in  a single clock time step, are called its \tit{nearest
  neighbors}, $nn$, the \tit{next-nearest neighbors},
$nnn$, are correspondingly defined and so on.

What we have defined up to now corresponds to the \tit{foliation} of
space-time into an aggregate of space-like slices. We now form the
union of these slices and define
\begin{equation}G:=\bigcup_t G(t)\end{equation}  
In our above mentioned papers (see in particular \cite{1}) we exploited
the fact that graphs carry a natural metric structure
\begin{equation}d(x_i,x_j):= \inf\{\text{length of paths, connecting
    $x_i$ and $x_j$}\}\end{equation}
where path length is the discrete number of edges of the path. This
defines a neighborhood structure on a graph.
\begin{equation}U_l(x_0):=\{\text{nodes $x_i$ with}\;
  d(x_0,x_i)\leq l\}\end{equation}

We now will transform $G$ into a \tit{partial ordered set} (\tit{poset}) by
introducing additional (\tit{causal}) bonds and relabeling the nodes.
From now on we denote the nodes in $G(t)$ by $x_i(t)$, that is, one and
the same node $x_i$ carries an additional time label $t\in
\Z\cdot\tau$, depending on the time slice $G(t)$ under discussion and
is denoted by $x_i(t)$. For each node, $x_i(t)$ we draw new edges to
the nodes $x_j(t+1)$ lying in $G(t+1)$, provided that $x_j(t)$ is a
$nn$ of $x_i(t)$ in $G(t)$ (including the node $x_i(t+1)$ itself!). For
convenience we usually drop the extra time element $\tau$.
\begin{defi}We call the edges lying in $G(t)$, that is the original
  edges of the (time dependent!) graph, the {\em spatial edges} (at
  time $t$), the edges which connect the neighbors in consecutive
  slices, $G(t),G(t+1)$, are dubbed {\em causal edges}. That is, an
  elementary {\em causal neighborhood} of, say, $x_i(t)$ consists of
  all the nodes, $x_j(t+1)$, in $G(t+1)$, with $x_j(t)$, having
  spatial distance, $d(x_i(t),x_j(t))\leq 1$, in $G(t)$ (that is, the
  node, $x_i(t+1)$ itself plus the nodes having distance one).
  \end{defi}
(It may be helpful to envisage the spatial edges as carrying a red
colour and the causal edges a green one).

We can now proceed by introducing the \tit{forward}- or
\tit{future cone}, \tit{backward}- or \tit{past cone}, respectively.
\begin{defi}To the forward cone of $x(t)$ belong those nodes,
  $y(t'),t'\geq t$, which can be connected by a causal edge sequence,
  $\gamma$, starting in $x(t)$. Such an admissible sequence consists
  of $(t'-t)$ elementary steps. An analogous definition holds for the
  members of the past cone. Given two nodes, $x(t),y(t')$ with $t'\geq
  t$, we can intersect the forward cone of $x(t)$ with the backward
  cone of $y(t')$ and get the corresponding double cone.
\end{defi}  
Remark: Note that the causal and metric relations are relatively
subtle as compared to, for instance, ordinary \tit{special
  relativity}, where we deal with one and the same topological space
structure for all times. In our space-time graph, $G$, the spatial
wiring is constantly changing on a microscopic scale, due to the
imposed local dynamical law. That is, two nodes may become nearest
neighbors in $G(t)$ while being far apart for earlier or later times
and vice versa. This can happen since bonds are permanently
annihilated and created.
\begin{conclusion}The above causal distance concept has already some
  of the crucial ingredients of the metric properties, known from
  \tit{general relativity}. Furthermore, it is of a markedly
  stochastic character.
\end{conclusion}

What we have said above, creates in a natural way some \tit{partial
  order} on the set of nodes. We do not want to reproduce all the
technical notions, which are presumably well known or can be found in
e.g. the papers of Sorkin et al, mentioned above, or in, say,
\cite{Birk} or \cite{Aigner}. In the definition of the partial ordered
set (poset), only the causal (green) bonds enter (whith their
(non)existence being a consequence of the respective (non)existence of
the spatial (red) bonds).
\begin{defi}We have $x_j(t')\geq x_i(t)\;,\;t'\geq t$ if the nodes can
  be connected  by a causal path, lying in the forward cone. The
  nodes, lying on a causal edge sequence, we call chains, sets of
  mutually space like nodes are called antichains.
\end{defi}
This order relation is clearly reflexive, antisymmetric and transitive. We
remark the following point. 

It trivially holds (by assumption) that $x_i(t')\geq x_i(t)$, that is,
for the same node at different times. This implies that for two nodes,
$x_i,x_j$ it follows
\begin{equation}x_j(t')\geq x_i(t)\;\Rightarrow\;x_j(t'')\geq
  x_i(t)\end{equation}
for all times, $t''\geq t'$, as we can continue the causal path from
$x_i(t)$ to $x_j(t')$ by the trivial path, $x_j(t')-x_j(t'')$.    
\section{The Geometric Coarse-Graining or Renormalisation Process}
\subsection{The General Picture}
One of our central hypotheses is it, to regard the ordinary space or
space-time as a medium having a complicated internal dynamical fine
structure, which is largely hidden on the ordinary macroscopic scales
due to the low level of (only mesoscopic) resolution of space-time
processes as compared to e.g. the Planck scale. The corresponding
process of \tit{coarse graining}, described in the following, may be
also called a \tit{geometric renormalisation}, in which the resolution
of the details of space-time is steadily scaled down to the level of
ordinary continuum physics. Some preliminary ideas of this
renormalsation process have already been been decribed in \cite{4} and
\cite{6}.

In the following we deal with a generic large network or graph, $G$,
as a typical representative of the members of the class $\{G(t)\}$,
described above. The individual renormalisation steps consist of the
following constructions.
\begin{itemize}
\item Starting from a given fixed graph, $G$, pick the (generic) \tit{cliques},
  $C_i$, in $G$, i.e. the subgraphs, forming maximal subsimplices or
  cliques in $G$ with their order lying in the above mentioned
  interval, $(r^0/2,r^0)$.
\item These cliques form the new nodes of the \tit{clique-graph},
  $G_{cl}$ of $G$. The corresponding new bonds are drawn between
  cliques, having a (sufficient degree of) overlap. Size, overlap and
  distribution of cliques in a generic (\tit{random}) graph have been
  analyzed in \cite{4}, for more details see the following subsection. 
\item That is, both \tit{marginal} cliques (if they do exist at all) and
  \tit{marginal} overlaps are deleted. In this respect a coarse-graining
  step includes also a certain \tit{purification} of the graph structure.
\end{itemize}
\begin{bem} What is considered to be a ``sufficient overlap'' depends of
course on the physical context and the general working philosophy. As
we noted above, a particular node will in general belong to several,
and in the case of densely entangled graphs to many, cliques. The
minimal possible overlap is given by a single common node. If, on the
other hand, the cliques on a certain level of coarse graining are
comparatively large, comprising, say, typically several hundred nodes,
it may be reasonable to neglect {\em marginal}, i.e. to small,
overlaps as physically irrelevant and define a sufficient degree of
overlap to consist of an appreciable fraction of the typical clique
order. Correspondingly, too small cliques, not lying in the above
introduced interval, $(r_0/2,r_0)$, are deleted (if they do exist at
all!, see the estimates in section \ref{cliques}). The numerical
effect of such choices will be studied in the following.
\end{bem}
\begin{defi}We call the graph, defined above, the (purified) clique graph,
  $G_{cl}$, constructed from the initial graph, $G$.
\end{defi}

It is an important question whether graphs and networks are connected,
that is, if there exists a path or edge sequence, connecting each pair
of vertices. This question becomes, a fortiori relevant, in the
following (sub)sections if the coarse-graining or renormalisation
steps are performed on a given fixed graph. The following lemma is
useful.
\begin{lemma}If $G$ is a connected graph, that is, each pair of
  vertices, $x,y$ can be connected by a finite path or edge sequence,
  depicted as $x=x_0-x_1-\cdots-x_n=y$, then the ordinary (unpurified)
  clique graph, $G_{cl}$ is again connected.
\end{lemma}
Proof: Let $x_0$ lie in a certain clique, $C_0$ and $y$ in a clique
$C_{n+1}$. By algorithmic construction (cf. \cite{4}), the vertices
$x_0-x_1,x_1-x_2,\ldots,x_{n-1}-x_n$ are lying in certain cliques
$C_1,\ldots,C_n$ with
\begin{equation}C_0\cap C_1\neq\emptyset\;,\;C_1\cap
  C_2\neq\emptyset\;\ldots,\;C_n\cap C_{n+1}\neq\emptyset\end{equation}
by construction ($x_i\in C_i\cap C_{i+1}$). Hence, each pair of
cliques, $C,C'$, can be connected by a finite sequence of pairwise
overlapping cliques. In other words, the ordinary clique graph is
again connected.\bewende 

This result is, for example, useful in cases where graphs are so
sparsely connected that, viewed in the random graph picture, there is
a non-zero probability that they are disconnected. The above
construction shows that at least the consecutive sequence of
unpurified clique graphs $G_0\to G_1\to G_2\to\cdots$ consists of
connected graphs, provided the initial graph, $G_0$, is connected,
with $G_{i+1}$ being the clique graph of $G_i$. On the other hand, if
we take instead the purified clique graph, in which only overlaps of a
certain degree are taken into account which are greater than some
prescribed value, it may happen that the clique graph is no longer
connected.

We want to repeat the above described coarse-graining process several
or perhaps many times (if necessary) without the necessity of
introducing new principles at each step of the construction. The
transition from a graph to its clique graph represents such a
\tit{universal principle}, which works on each level of the
renormalisation process.  In the end we hope to arrive at a
(quasi-)continuous manifold, displaying, under appropriate
magnification, an intricate internal fine structure. This should (or
rather, can only expected to) happen if the original network has been
in a (quasi-)\tit{critical} state as will be described in the
following (see in particular section 8).

On each level of coarse-graining, that is, after each renormalisation
step, labelled by $l\in\Z$, we get, as in the block spin approach to
critical phenomena, a new level set of cliques or lumps,$ C^l_i$, ($i$
labelling the cliques on renormalisation level $l$), consisting on
their sides of $(l-1)$-cliques which are the $l$-nodes of level $l$,
starting from the level $l=0$ with $G=:G_0$. That is, we have
\begin{equation}C^l_j=\bigcup_{i\in j}
  C_i^{(l-1)}\;,\;C_i^{(l-1)}=\bigcup_{k\in i}
  C_k^{(l-2)}\;\text{etc.}\end{equation}
($i\in j$ denoting the $(l-1)$-cliques, belonging, as meta nodes,
 to the $l$-clique, $C_j$).
These cliques form the meta nodes in the next step.
\begin{defi}The cliques, $C_i^0$, of $G=:G_0$ are called zero-cliques. They
  become the one-nodes, $x_i^1$, of level one, i.e. of $G_1$. The one-cliques,
  $C_i^1$, are the cliques in $G_1$. They become the 2-nodes, $x_i^2$,
  of $G_2$ etc.  Correspondingly, we label the other structural
  elements, for example, 1-edges, 2-edges or the distance functions,
  $d_l(x_i^l,x_j^l)$. These higher-level nodes and edges are also
  called meta-nodes, -edges, respectively.
\end{defi}

The following figure shows how the (meta) nodes and bonds form in two
consecutive steps. In this example and in the selected subgraph under
discussion the cliques on level $0$ are triangles. Some of them have a
common bond but all of them are hanging together via a common
(central) node. In this example we draw a bond on level $1$
if the cliques of level $0$ have at least one node in common.
\begin{figure}[h]
\centerline{\epsfig{file=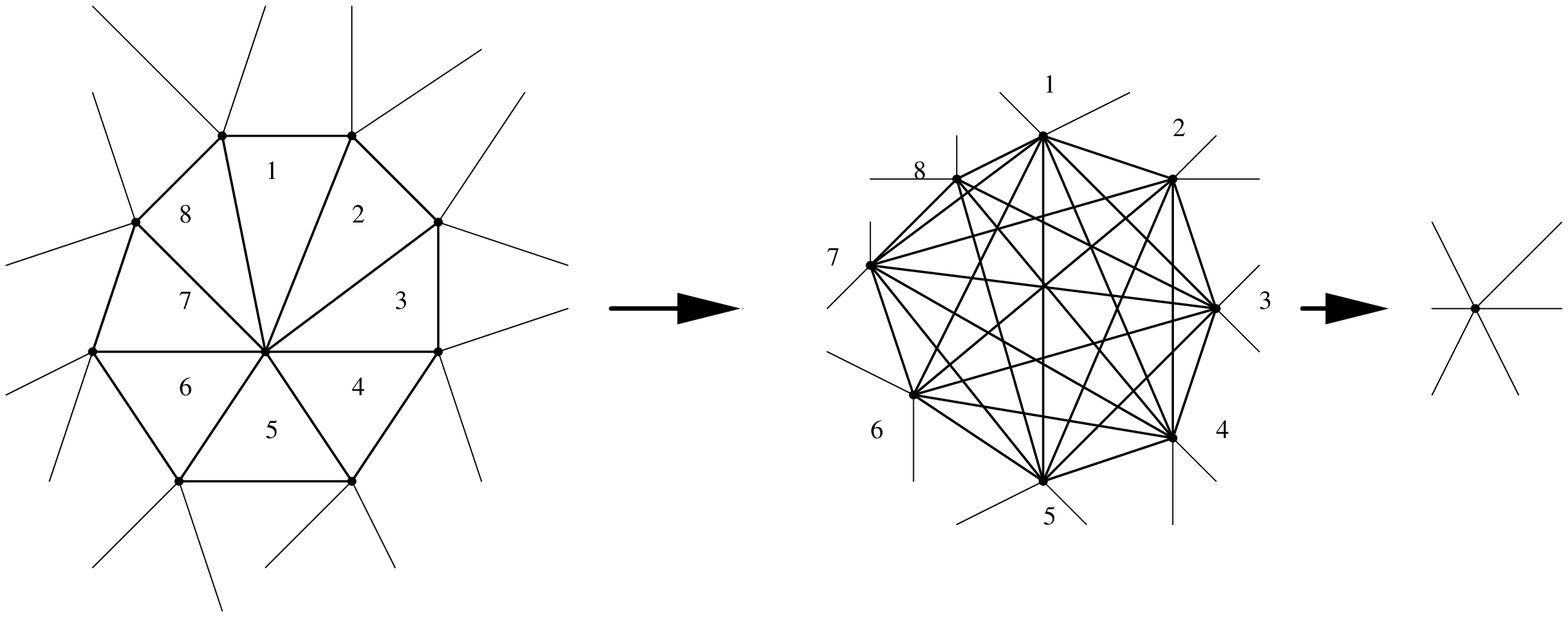,width=10cm,height=4cm,angle=0}}
\caption{}
\end{figure}

\begin{bem}The picture may lead to the wrong impression that the
  network becomes sparser after each step. Quite to the contrary, the
  number of cliques in $G_{cl}$ may be much larger than the number of
  nodes in the original graph, $G$ (cf. the table in section
  \ref{cliques}). This happens if there is an appreciable overlap
  among the occurring cliques, that is, a given node may belong to
  many different cliques. On the other hand, after several
  renormalisation steps, the picture becomes stable in the generic
  case (see the following subsection).
\end{bem}

The above illustration can be understood in two different ways. On the
one hand, read from left to right, the resolution of space appears to
be reduced. The cluster of cliques on the left happens to be
contracted to a single node of the next level. On the other hand,
according to our working philosophy, we can regard the node on the
right as still containing the structure on the left, which could, in
principle, be recovered when increasing the resolving power of our
space-time microscope, i.e., by increasing e.g. the energy. This is
expressed in the following picture (where for the sake of graphical
clarity, the mutual overlaps of the occurring cliques of the same
level is not represented!).
\begin{figure}[h]
\centerline{\epsfig{file=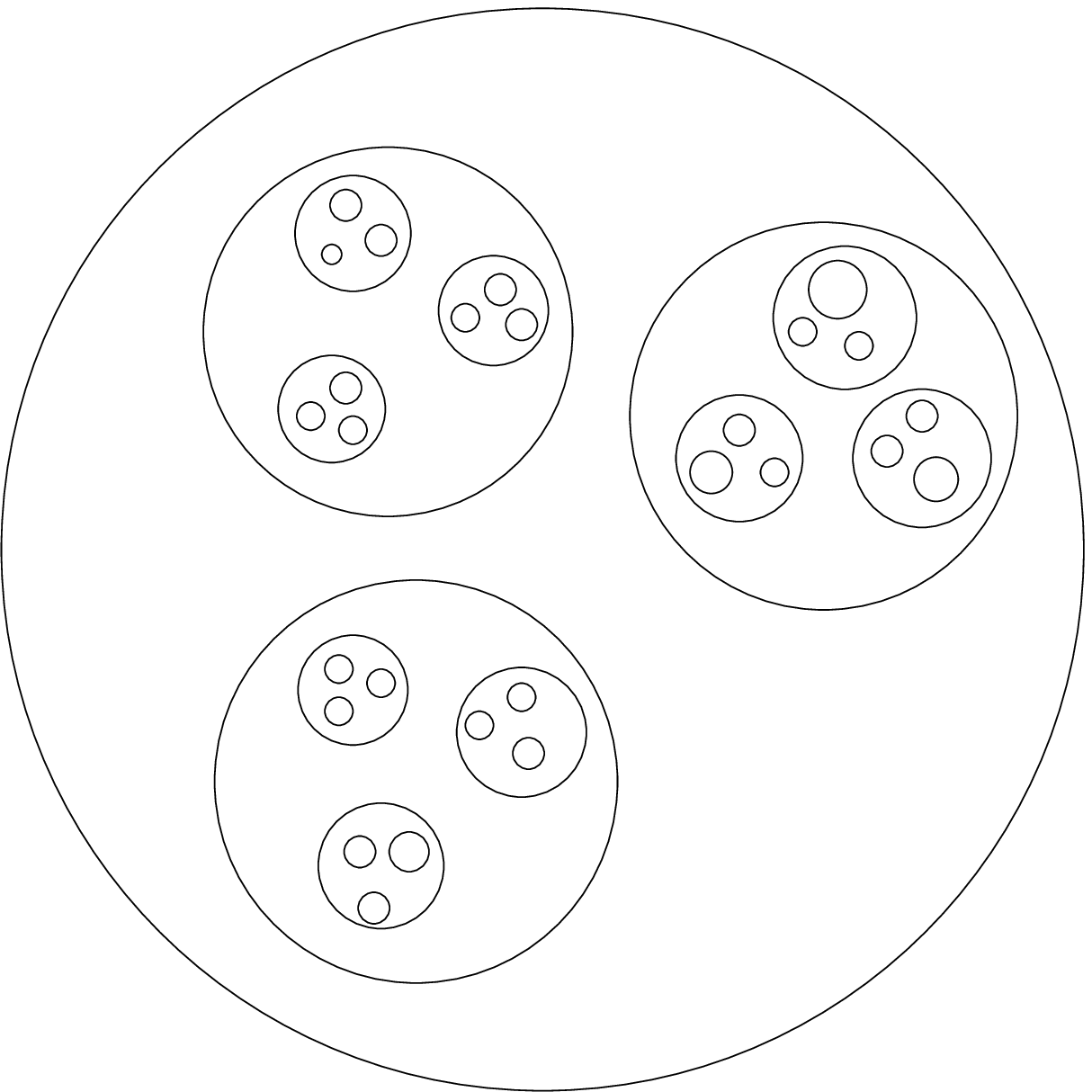,width=6cm,height=6cm,angle=0}}
\caption{}
\end{figure}
Understood in this latter sense we call these space-time points also
\tit{lumps}, that is, we regard them as objects, having an inner
structure. Different aspects of this structure emerge on the
respective scale of resolution or magnification. We provided arguments
in \cite{Quantum} that in our view even quantum theory is just such an
emergent aspect which shows up at the typical quantum scale.

We want to briefly mention the possibly far-reaching interplay on the
higher levels of coarse-graining between these deleted, too marginal,
overlaps and the more \tit{local} wiring stemming from the
non-marginal overlaps. We discussed this point at length in
\cite{Quantum}. We argued there that these deleted meta bonds are
responsible for the translocal behavior of quantum theory. In the
following we are, however, chiefly be concerned with the emergence of
smooth and local behavior, leading, hopefully, to (quasi)classical
space-time structures).
\subsection{ \label{ana} The Analytic and Numerical Results}
We begin this subsection with a general remark concerning the
character of our approximations.
\begin{bem}As the individual terms in our combinatorial expressions
  are typically either extremely large or small and are frequently, as
  in statistical mechanics, very sensitive to the given range of
  parameters, it is a quite delicate matter to make safe estimates.
  Among other things we usually have to take logarithms and compare
  them. That is, if for example $\log(a)\gg \log(b)$, we sometimes
  choose to neglect $\log(b)$ in a contribution like $\log(a)+\log(b)$
  in the further calculations. For the original expression this may
  have the effect that we replace $b\cdot a$ by $a$. To give an
  example; we sometimes approximate $10^2\cdot 10^{100}$ by
  $10^{100}$. Otherwise we had to take into account a lot of only
  marginal contributions which would make the calculations rather
  cumbersome. On the other hand, this is of course only justified, if
  we are only interested in qualitative results and provided that the
  final result is insensitive to such an approximation.
\end{bem}
We made more detailed remarks in \cite{4} formulas (62) ff., where we
discussed the approximation of e.g. binomial coefficients and their logarithms.

We have seen that the cliques in a large generic random graph of \tit{order}
$n$ and \tit{bond probability} $p$ are with high probability
concentrated in the interval $(r_0/2,r_0)$ with respect to their
order, $r$, with
\begin{equation}r_0\approx 2\log(n)/\log(p^{-1})+ O(\log\log(n))\end{equation}
and with the expectation of $r$-cliques
\begin{equation}\label{Zr}   \langle Z_r
  \rangle=\binom{n}{r}\cdot(1-p^r)^{n-r}\cdot
  p^{\binom{r}{2}}\end{equation}
We can test our general working philosophy concerning the effects of
coarse graining and renormalisation by analytically and numerically
calculating various properties of the \tit{clique graph} of a generic
random graph. These calculations become increasingly intricate with
increasing complexity of the asked questions. Some of the analysis has
already been done in e.g. sect. 4.2 of \cite{4} (called `\tit{The
  Unfolded Epoch}') to which we refer the reader for more
technical details.

The meta nodes of the clique graph, $G_{cl}$, are the cliques of $G$.
The meta bonds in $G_{cl}$ are given by the overlap of cliques in $G$.
As we want, on physical grounds, to ignore marginal, that is, too
small overlaps, it is important to calculate the expected number of
$r'$-cliques, $\langle N(C_0;r',l)\rangle$, having an overlap of order
$l$ with a given fixed $r$-clique, $C_0$ with both $r$ and $r'$ lying
in the above interval of generic cliques.

In \cite{4} sect. 4.2 we derived the following formula for this
stochastic quantity ($C_0$ being a fixed $r$-clique):
\begin{equation}\label{N}  \langle N(C_0;r',l)\rangle=\frac{\binom{r}{l}\cdot
 \binom{n-r}{r'-l}\cdot
      p^{\binom{r'}{2}-\binom{l}{2}}}{(1-p^r)^{n-r}\cdot
      }\cdot P_{r',l}
\end{equation}
with $P_{r',l}$ a lengthy combinatorial expression (formula (69) in
\cite{4}) which we can neglect for the parameters $n,r,r'$, chosen
by us, that is $n\ggg r,r'\gg l$. That is, in this regime we 
approximated  $P_{r',l}$ by one. It can however not be neglected if this
assumption is violated!.

After some manipulations we arrive at the following approximative
formula (\cite{4}, formula (74)), where we choose, for convenience,
$r'=r$, as we are at the moment only interested in qualitative or
generic results:
\begin{multline}\log\langle
  N(C_0;r',l)\rangle\approx(r'-l)\cdot(\log(n)-\log(r'-l))+1/2\cdot
 (r')^2\cdot\log(p)\\\approx \log\langle Z_{r'}\rangle-l\cdot\log(n)-r'\log(r')  \end{multline}
with 
\begin{equation}\log\langle Z_r\rangle\approx
  r\log(n)+1/2\cdot r^2\cdot\log(p)\end{equation}
for this range of parameters (cf. also \cite{4} formula (60)ff).

The total expected number of $r'$-cliques, having an overlap $l\geq
l_0$ with a given $r$-clique is
\begin{equation}\sum_{l\geq l_0}\langle N(C_0;r',l)\rangle    \end{equation}
(the admissible $l$'s being bounded by the minimum of $r$ and
$r'$). For $l=0$ we get the expected number of $r'$-cliques, having
zero overlap with the given fixed $r$-clique, $C_0$, that is we have
approximately (remember our simplifying assumption $r=r'$): 
\begin{equation}\log\langle N(C_0;r',l=0)\rangle\approx\log\langle Z_{r'}\rangle-r'\cdot\log(r')
\end{equation}

As $n$ is so large, the total number of $r'$-cliques, having overlap
$l\geq l_0$ with $C_0$ can be approximated by the number of cliques
fullfilling the lower bound $l_0$. On the other hand, the total number of
expected generic cliques, $N_{cl}$, in the random graph, $G$, that is,
the cliques with order lying in the respective interval $(r_0/2,r_0)$
is roughly
\begin{equation}N_{cl}\approx r_0/2\cdot \langle Z_{\bar{r}}\rangle      \end{equation}
with $\bar{r}$ an appropriate value in the above interval (this replacement
can be made as the numerical values in this interval behave relatively
uniformly). We define the \tit{local group} of a generic clique as the
set of generic cliques, having non-marginal overlap with the fixed
given clique. From the above reasoning we can now infer the
following important conclusion
\begin{conclusion}
\begin{equation}\langle N_{loc.gr.}\rangle\approx N_{cl}/(n^{l_0}\cdot\bar{r}^{\bar{r}})
 \end{equation}    
with $n$ the number of nodes in the graph, $G$, $N_{cl}$ the
number of generic cliques in the corresponding clique graph, $l_0$ the
degree of overlap of the generic cliques, $\bar{r}$ some appropriate
value in the interval $[r_0/2,r_0]$, $n\gg r,r'\gg l_0$ being assumed
(where the second $\gg$ is not so pronounced as the first one; $n$ is
usually gigantic compared to the clique size $r$!).
\end{conclusion}
Such estimates are central in the following as they provide information
about the local structure of the clique graph. 

From the above formulas and numerical results we can now infer
interesting properties of the clique graph of a typical graph of
order, $n$, and bond probability, $p$. The expected order of the \tit{local
  group} in the clique graph is, by the same token, the \tit{average
  vertex degree} in the clique graph. That is
\begin{equation}\langle v_{cl}\rangle\approx N_{cl}/(n^{l_0}\cdot\bar{r}^{\bar{r}})
 \end{equation}    
From this we can immediately infer the bond probability of the clique graph:
\begin{equation}p_{cl}=\langle v_{cl}\rangle/(N_{cl}-1)\approx \langle
  N_{loc.gr.}\rangle/N_{cl}\approx n^{-l_0}\cdot \bar{r}^{-\bar{r}}\end{equation}
and see that it is already considerably smaller than the bond
probability of the underlying microscopic graph we started from which,
in our numerical example, was assumed to be of order one.

We take our above numerical example, $n=10^{100}\,,\,p=0.7$ which
implies $r_0=1291$ and assume that an appreciable overlap for generic
cliques should be of the order of, say, 50 nodes. As typical clique
size we take $\bar{r}=r_0/2$ (remember that we are at the moment only
interested in qualitative results). The clique graph has roughly
$N_{cl}\approx 10^{10^4}$ generic cliques, that is, meta nodes of the
first level. With the bond probability in the clique graph,
$p_{cl}\approx 10^{-7\cdot 10^3}$, we now can calculate the
distribution and order of cliques of the first level, that is,
cliques of cliques. This provides important information about the near
order of the clique graph and the effects of the renormalisation
steps.

As the order of these cliques of the first level turns out to be
already quite small, it is reasonable to avoid our approximative
formulas and determine the respective clique number, $r_0$, by
explicitly calculating the number where $\langle Z_r\rangle$ drops
from a very large number to effectively zero. The result shows, that
for $overlap=50$ of the original cliques (of the zero level), the
cliques of the next higher level comprise only very few cliques of the
zero level. That is, the near-order of $G_1:=G_{cl}$ is already much
coarser or less erratic as compared to the near order in the original
graph.  The results are described in the following observation.
\begin{ob}For the above numerical parameters we get a typical clique
  size on the first coarse grained level of order $r=2$ or $3$ and an
  expected number of cliques of the first level of the order of
  $\log(N_{cl})=10^4$ (which is comparable to the number of cliques of
  the zero level!).
\end{ob}
We can control the sensitivity of our results to the chosen degree of
overlap. We see below that the results do not depend critically on the
numerical details as long as the parameters are roughly of the same
order. For e.g.  $overlap=30$ we get, performing the corresponding
calculations, the following result.
\begin{ob}For clique-$overlap=30$ the clique size on the first level increases
  sligtly to a value of $r_0=4$.
\end{ob}

In the following we present some more characteristics of the clique
graph with overlap 50.
\begin{itemize}
\item average vertex degree $\approx 10^{(10^4-7\cdot 10^3+3)}$
\item expected number of bonds $=\langle v_{cl}\rangle\cdot
  0.5\cdot\langle n_{cl}\rangle\approx 0.5\cdot10^{5\cdot
    10^3+10^4+6}$
\end{itemize}
An important question is whether the (purified) clique graph, $G_1$,
is still connected. In \cite{4} we gave the threshold value for the
corresponding $p^*(n)$, which is
\begin{equation}p^*(N_{cl})=\log(N_{cl})/N_{cl}\approx
  10^4/10^{10^4}=10^{-(10^4-4)}\ll p_{cl}\approx 10^{-7\cdot 10^3}\end{equation}
that is,
\begin{equation}p_{cl}/p^*(n)\approx 10^{3\cdot 10^3}    \end{equation}
\begin{conclusion}For the numerical data we employed the web of lumps
  is {\em almost surely} connected. On the other hand, after one
  renormalisation step, the purified net of cliques is much sparser
  connected than the initial microscopic net.
\end{conclusion}

Summing up, what we have accomplished so far in this subsection, we
have the following row of graph characteristics for the particular set
of numerical parameters we employed.
\begin{itemize}
\item $l=0$: number of nodes $n_0=10^{100}$, bond probability $p_0=0.7$,
clique number $r_0=1291$.
\item $l=1$: $n_1\approx 10^{10^4}$, $p_1\approx 10^{-7\cdot 10^3}$,
  $r_1= 3$, $\langle vertex\; degree\rangle\approx 10^{0.3\cdot 10^4}$
\end{itemize}
The respective values were calculated by using the following
approximative formulas:
\begin{equation}p_1=n_0^{-l_0}\cdot
  \bar{r}^{-\bar{r}}\;,\,l_0=50\;,\;n_1\approx r_0/2\cdot\langle
  Z_{\bar{r}}^{(0)}\rangle \end{equation}
($\langle Z_r^{(0)}\rangle$  the distribution function of cliques in
the initial graph, $G_0$, $\bar{r}$ some average or typical value).

The expected order of cliques on level $1$ is only $2$ or $3$.  That
is, taking the next step from level $1$ to level $2$ we may assume an
overlap $l_1=1$, i.e., we may take the ordinary clique graph. With
this value we can calculate the corresponding characteristics of
$G_2$, the graph having as nodes cliques, consisting of nodes of level
$1$. Before we proceed with the numerical estimates we first have to
check whether the approximations we have made above are still valid
for this new regime of parameters!

Now, $r,r',l$ are both very small and of comparable size. That is, our
above approximative formulas are no longer valid. On the other hand,
for $r,r',l$ near one, it becomes possible to evaluate the
combinatorial expressions directly. For the expected number of nodes
on level 2, that is, expected number of cliques on level 1, we insert
our parameters into the formula for $\langle Z^{(1)}_r$ (cf. formula
(\ref{Zr})) and get an approximate value, $n_2\approx 10^{10^4}$
(which is of the same order as $n_1$!).

The calculation of the vertex degree, that is, $\langle
N^{(1)}(C_0,r',l)\rangle$ with e.g. $r'=2,l=1$, is numerically more
delicate since now we have to take into account also the term
$P_{r',l}$ in formula (\ref{N}), we up to now approximated by
one. Furthermore we now face the problem of having to deal with
small differences of extremely large numbers in the various occurring
expressions and/or factors which are extremely small or large and tend
to cancel each other. 

Fortunately, there is a more direct way to get sufficiently precise
results in this regime. We saw that typical cliques in $G_1$ are of
order two or three. The assumed overlap is $l=1$. We can hence infer
that the expected number of cliques, overlapping with a fixed given
clique, $C_0$, is roughly the same as the number of nodes, being
connected with one of the nodes of $C_0$. We conclude that
\begin{equation}p_2=\langle N^{(1)}(C_0,r',l)\rangle\approx p_1\approx
  10^{-7\cdot 10^3}\end{equation}
With these values for $n_2,p_2$, we can calculate $r_2$ and again get $r_2=3$. 
We hence have for $l=2$:
\begin{itemize}
\item $n_2\approx 10^{10^4}$ (number of cliques of level
  $1$)
\item $p_2\approx p_1\approx 10^{-7\cdot 10^3}$, $r_2=3$
\end{itemize}
For the following levels the parameters are now stable and the same as
for level two.
\begin{conclusion}We see that after only two steps we have arrived at
  a coarse grained graph with a large number of nodes, a very small
  bond probability and small cliques, which shows that the geometric
  near- and far-order has unfolded. We further conclude that the
  following renormalsation steps would no longer alter appreciably the
  graph characteristics calculated above for the levels $G_1,G_2$.
  That is, at least as far as these particular graph properties are
  concerned, we have already reached a quasi-stable regime, so that
  the assumption of the existence of fixed phases or attractors does
  not seem too far-fetched. We can also infer that all the graphs are
  almost shurely connected.
  
  On the other hand, we do not expect that a smooth limit manifold,
  having e.g. a fixed integer dimension, does emerge quasi
  automatically in the pure random graph framework.  A further
  important ingredient will be the action of some appropriately chosen
  local law as we have introduced it above.  (see the corresponding
  discussions in our mentioned prior work).
\end{conclusion}
\section{Fixed Point Behavior}
Starting from a sufficiently large network or graph, $G=G_0$, and
performing the consecutive steps, described above, denoting the
transition from $G_l$ to $G_{l+1}$, i.e. from a graph to its
(purified) clique graph, by $R$ (standing for \tit{renormalisation}),
we have
\begin{equation}R:G_l\to G_{l+1}\;,\;G_l=R^lG_0\;,\;R^l=R\cdots
  R\;\text{($l$-times)}\end{equation}

The philosophy of the renormalisation group is, that initial systems,
lying on the \tit{critical submanifold}, approach a \tit{fixed point}
under $R^l$ for $l\to\infty$. In statistical mechanics the limit
systems represent rather a \tit{limit phase}, i.e. a statistical system
with the finer details still fluctuating. In the same sense we can at
best hope that our presumed limit network of lumps represents a
similar limit phase, that is, a network which is only invariant and
homogeneous on a larger scale of resolution, while the fine structure
is still constantly changing.

The geometric concepts, which have to be further clarified, are the
notions of \tit{geometric (fixed) phase} and \tit{critical network
  state}. We want to emphasize that we cannot expect that these
characterisations will be a simple task. Quite to the contrary, both
concepts represent subtle and delicate properties. In general, the
emerging array of lumps will not fit automatically into something
which does resemble a smooth macroscopic manifold, having for example
a well-defined and integer (macroscopic) dimension (among other
things).  Possible obstacles are already well-known on the much
simpler level of simplicial complexes. In order that such a complex
has the chance to approximate a manifold, a variety of subtle incident
relation between the occurring individual simplices have to be
fulfilled (see e.g.  \cite{Stillwell}).

In our context these relations on the more coarse-grained scales will
depend on the appropriate choice of the microscopic local dynamical
laws on the Planck scale we started from. Experience with
complex systems in general and cellular automata in particular tells
us, that the class of appropriate laws will be a very small and
peculiar set in the space of possible interaction laws. See the
corresponding findings in the regime of \tit{selforganized
  criticality} (\cite{Bak}), the catchword being \tit{complexity at
  the edge of chaos}.

In other words, as the whole approach appears to be relatively new and
the task formidable, we will make what are perhaps only some first
steps towards a solution of these problems. In a first step we will
convince ourselves that the renormalisation procedure described by us
does not lead to nonsensical results (we have already previously seen
that some gross characteristics of the network seem to become stable
after only a few renormalisation steps). We show that there do exist
examples of graphs which display fixed point or fixed phase behavior
in a more microscopic sense.  These graphs are however simple and very
regular and are not meant to represent possible examples of networks,
underlying our continuum space-time. They rather serve at best as
illustrative toy models.

In the following section we then introduce a geometrical core concept
designed to classify such irregular network structures, i.e. the
notion of \tit{graph dimension}. We show, how it behaves under our
renormalisation process. The corresponding analytic results indicate
what kind of \tit{critical behavior} is presumably needed to have a
physically reasonable limit behavior.

We illustrate our framework with the help of some simple examples (see
also the following figure 5). Note that in the following examples the
minimal admissible
clique overlap is assumed to be one common node! \\[0.3cm]
1)\tit{The graph $\Z^2$}:\\[0.3cm]
The set of nodes are parametrized as $V=\{(i,j),\,i,j\in\Z\}$. Edges
are drawn between the following nodes:
\begin{equation}(i,j),(i',j')\;\text{with}\;|i'-i|+|j'-j|=1\end{equation}
We determine the cliques at the various levels, given by $G_l$ (see
also the following figure).\\[0.2cm]
$G_0$) A node, $(i,j)$, belongs to the following $0$-cliques:
\begin{equation}\{(i+1,j),(i,j)\}\;\{(i,j+1),(i,j)\}\end{equation}
and $+$ replaced by $-$. That is, the order of the $0$-cliques is $2$,
the diameter (that is, the maximal distance between two nodes) is $1$, the maximal mutual overlap is $1$.\\[0.2cm]
$G_1$) A $0$-node, $(i,j)$ belongs to the following $1$-cliques
\begin{equation}\{(i,j),(i\pm1,j)(i,j\pm1)\}\end{equation}
and the cliques, formed around the $nn$-nodes of $(i,j)$. The order
relative to $G_0$ is $5$, the diameter is $2$, maximal overlap is
$2$.\\[0.2cm]
$G_2$) The order of $2$-cliques relative to $G_0$ is $13$, diameter is
$3$, maximal overlap is $8$.\\[0.3cm]
Remark: Note that the above values of order and diameter refer to the
start graph $G_0$.\vspace{0.3cm}

With increasing $l$, the maximal overlap becomes large, due to the
particular structure of the graph, $\Z^2$. One sees that for large $l$
the hierarchical structure of the corresponding tower of graphs,
$G_l$, becomes very dense and entangled, a feature one would also
expect from something like a \tit{continuum}.

On the other side, it is instructive, to perform also the above
mentioned rescaling and compare the various levels at the same scale,
viz., inspect the pure graph structure. This will make explicit the
fixed point behavior, we are particularly interested in.\\[0.2cm]
$G_0\to G_1$) The $1$-nodes of $G_1$ (i.e. the $0$-cliques) we
represent by the midpoints of the edges of the start graph,
$G_0:=\Z^2$. Four of these $0$-cliques meet at a common node, $(i,j)$,
say. We represent the $1$-edges as the line segments, connecting these
midpoints. This yields a new, rotated lattice (pus two extra
diagonal edges).\\[0.2cm]
$G_1\to G_2$) These four $1$-nodes (the $1$-cliques) form now the
$2$-nodes. They form a simplex having $6$ $1$-edges. We inscribe these
$2$-nodes in $G_0$ by placing them in the centers of the $1$-cliques,
that is the original lattice points of $G_0$. We draw a $2$-edge if
two of these $1$-cliques have a common $1$-node (that is, a
$0$-clique!). We can convince ourselves that the emerging graph, $G_2$
is isomorphic to the start graph, $G_0$. We hence make the interesting
observation:
\begin{ob}Starting from $G_0=\Z^2$, we see that $G_2$ is
  combinatorially isomorphic to $G_0$, meaning that there exists an
  invertible map, $\Phi:G_0\to G_2$, mapping nodes on nodes and
  bonds on bonds and preserving the combinatorial structure in the
  following way (with $e_{ij}$ an edge of $G_0$)
\begin{equation}e_{ij}\in E(G_0)\leftrightarrow
  \Phi(e_{ij})\;\text{connects}\;\Phi(x_i),\Phi(x_j)\end{equation}
The same holds for $G_1,G_3$ etc.
\end{ob}
\begin{conclusion}The sequence of graphs, $G_0,G_1,G_2,\ldots$,
 decomposes in exactly two sets of isomorphic graphs, 
\begin{equation}\{G_0,G_2,\ldots\}\;,\;\{G_1,G_3,\ldots\}  \end{equation}
under the renormalisation group
\begin{equation}\mcal{R}:=\{R^i\}\;,\;R^i:G_0\to G_i\;,\;R^2:G_i\to G_{i+2}   \end{equation}
\end{conclusion}
\begin{koro}A corresponding observation can be made for a general
  lattice, $\Z^n$.
\end{koro}
\begin{figure}[h]
\centerline{\epsfig{file=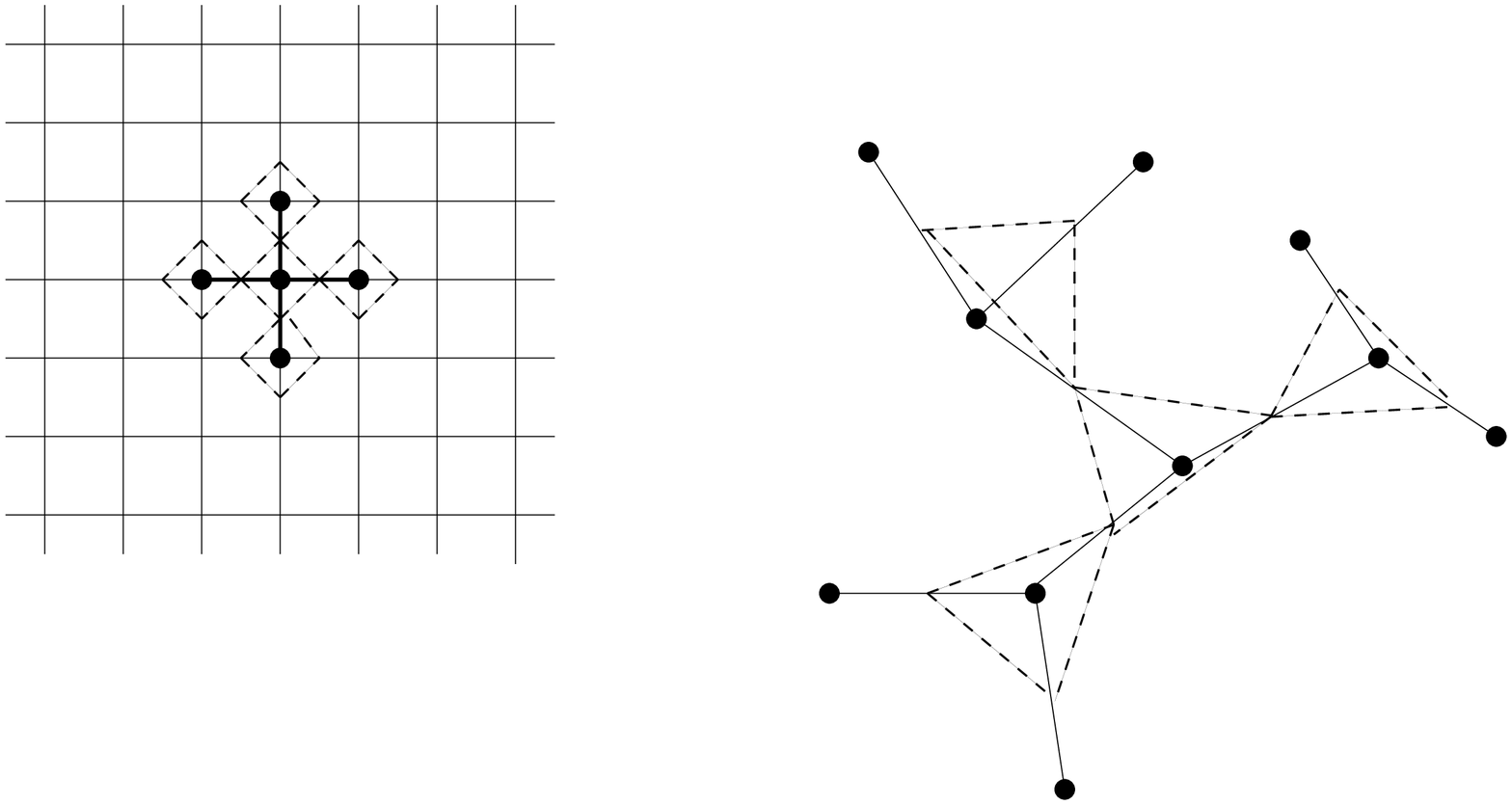,width=10cm,height=6cm,angle=0}}
\caption{}
\end{figure}
2)\tit{The trivalent infinite tree}:\\[0.3cm]
In order to get a better feeling for what can happen, we study some
more elementary examples. Let us take an infinite trivalent tree. The
$0$-cliques are again $2$-sets of vertices or line segments,
connecting $nn$. The graph, $G_1$, is again represented by
connecting the midpoints of these line segments. The resulting
$1$-cliques are $3$-sets or triangles. Taking them as the $2$-nodes of
$G_2$, we see that $G_2$ is again isomorphic to $G_0$ as in the $\Z^2$-case.
\begin{ob}For a trivalent infinite tree, the sequence of graphs,
  $G_0,G_1,G_2,\ldots$, decomposes into two subsets. The
  situation is the same as for the preceding example.
\end{ob}
3)\tit{The triangulated $\R^2$}:\\[0.3cm]
We introduce another simple example. We triangulate $\R^2$ by using
the above lattice, $\Z^2$, and complement it by drawing the diagonals,
pointing from $(i,j)$ to $(i+1,j+1)$. The $0$-cliques are these
triangles. Without a purification, bonds in the graph, $G_1$, are
drawn if two $0$-cliques meet at a common node or $0$-edge. The
emerging $1$-cliques have the shape of hexagons, i.e. they are
$6$-simplices. Repeating this process, one sees that $G_2$ is
isomorphic to $G_1$.
\begin{ob}In the case of the above triangulation of $\R^2$, we have a
  start graph, $G_0$, while all the graphs, $G_1,G_2,\ldots$, are
  isomorphic. In other words, we now have a fixed point of the
  renormalisation group.
\end{ob} 
\begin{conclusion}We have seen that there exist examples in the category of
  graphs which display phenomena like invariant sets or fixed points
  under our geometric renormalisation group.
\end{conclusion}
These observations open up interesting vistas. While we have not yet
shown that the above invariant sets or fixed points have the character
of \tit{attracting sets}, that is, whether there exist large
\tit{basins of attraction} in the category of graphs under the
repeated application of the map, $R$, we strongly surmise that this is
true.  Furthermore, the concept of \tit{selfsimilarity} suggests
itself (see also the next section), a notion we have already
introduced and studied in \cite{1}, to construct graphs with
\tit{fractal dimension}.
\section{Graph Dimension under the Renormalisation Group}
We repeatedly mentioned the possibility of \tit{geometric} or
\tit{topological phase transitions} in evolving networks of the kind
we are having in mind. In \cite{1} we developed and studied the
concept of \tit{graph dimension} in quite some detail. We concluded
that, from the physical point of view, the number of nodes which can
be reached by, say, $l$ steps starting from a given node, is an
important characteristic as is its limiting and scaling behavior as a
function of $l$. This is the crucial and \tit{intrinsic} property,
which underlies implicitly most of the calculations in the physics of
phase transitions and many other phenomena, which are triggered by the
collective interaction of many constituents. Its true significance is
however frequently hidden as the reasoning is usually performed by
using the properties of the
embedding space (viz., its ordinary dimension). \\[0.3cm]
Remark: We learned recently that such growth properties are also
important characteristics in geometric group theory and related
subjects in pure mathematics (see e.g. \cite{Harpe},\cite{Bartholdi}
or \cite{Grigorchuk}).\vspace{0.3cm}

We will investigate the behavior of this quantity under the
application of our renormalisation group. In \cite{1} we introduced
the two variants, defined below. They are not strictly equivalent but
coincide in the more regular situations. In the following, for the
sake of brevity, we only use the first notion.
\begin{defi}[Internal Scaling Dimension] 
  Let $x$ be an arbitrary node of $G$. Let $\#(U_n(x))$ denote
  the number of nodes in $U_n(x)$.We consider the sequence of
  real numbers $D_n(x):= \frac{\ln(\#(U_n(x))}{\ln(n)}$. We say
  $\underline{D}_S(x):= \liminf_{n \rightarrow \infty} D_n(x)$ is the
  {\em lower} and $\overline{D}_S(x):= \limsup_{n \rightarrow \infty}
  D_n(x)$ the {\em upper internal scaling dimension} of G starting
  from $x$. If $\underline{D}_S(x)= \overline{D}_S(x)=: D_S(x)$ we say
  $G$ has internal scaling dimension $D_S(x)$ starting from $x$.
  Finally, if $D_S(x)= D_S$ $\forall x$, we simply say $G$ has {\em
    internal scaling dimension $D_S$}.
\end{defi}
\begin{defi}[Connectivity Dimension] 
  Let $x$ again be an arbitrary node of $G$. Let $\#(\partial
  U_n(x))$ denot the number of nodes in the boundary of $U_n(x)$. We
  set $\tilde{D}_n(x) := \frac{\ln(\#(\partial U_n(x))}{\ln(n)} +1$ and
  define $\underline{D}_C(x) := \liminf_{n \rightarrow \infty}
  \tilde{D}_n(x)$ as the {\em lower} and $\overline{D}_C(x) :=
  \limsup_{n \rightarrow \infty} \tilde{D}_n(x)$ as the {\em upper
    connectivity dimension}.  If lower and upper dimension coincide,
  we say $G$ has {\em connectivity dimension} $D_C(x) :=
  \overline{D}_C(x) = \underline{D}_C(x)$ {\em starting from} $x$. If
  $D_C(x) = D_C$ for all $x$ we call $D_C$ simply the {\em
    connectivity dimension} of $G$.
\end{defi}
Remark: The above does not imply, that this notion is the only
relevant topological characteristic of large networks. It clearly is
not sufficient, to describe \tit{all} of the mesoscopic or macroscopic
properties, but we think it is, as in the continuum, a very important
concept.\vspace{0.3cm}

We already proved in \cite{1} that this kind of dimension is stable
under a variety of transformations, in particular under \tit{local}
ones. In section 5.2.5 of \cite{1} we showed that, in order to change
the dimension of a graph, we have to introduce long-range effects or
interactions. This reminds one of the behavior of \tit{critical
  systems}.

We now compare the dimension of a graph, $G$, with the dimension of
its clique graph, $G_{cl}$, where, for the time being, we take the
clique graph in its original meaning. That is, we draw a bond if two
cliques have a non-void overlap of arbitrary size.

Let us assume, for convenience, that $G$ has the scaling dimension,
$D$, that is, for every node, $x_0$, we have
\begin{equation}\lim_l \ln\,(\#(U_l))/\ln \,l=D\end{equation}
Furthermore, we assume for simplicity that the node degree of $G$ is
globally bounded, i.e.
\begin{equation}v_i\leq v<\infty\;\text{for all}\;x_i\end{equation}

We choose a fixed node, $x_0$, lying in a fixed clique, $C_0$. We have
to calculate the number of $1$-nodes, that is, the number of
$0$-cliques, $\#(U_l^{cl}(C_0))$, lying in $U_l^{cl}(C_0)$ with the
distance, $d_1$ now measured in the clique graph, $G_{cl}$. That is, a
clique, $C_l$, lies in $U_l^{cl}(C_0)$ if $C_0$ and $C_l$ can be
connected by a sequence of $l'$ cliques with $l'\leq l$ so that two
consecutive cliques have a non-zero overlap.  For each $0$-node,
$x_l'$, lying in some $C_{l'}$ with $d_1(C_0,C_{l'})\leq l$, we can
estimate the distance to the node $x_0$ in $C_0$. There exists, by
definition, a sequence of overlapping cliques,
\begin{equation}C_o,C_1,\ldots,C_{l'}\;,\;l'\leq l \end{equation}
For two neighboring cliques, $C_i,C_j$, we have
\begin{equation}d_0(x_i,x_j)\leq 2\;,\;x_i\in C_i,x_j\in C_j  \end{equation}

For each intermediate consecutive pair of cliques we need one step (a
bond from a node in the overlap $C_{i-1}\cap C_i$ to a node in
$C_i\cap C_{i+1}$), for the initial and final pair we need at most two
steps, we hence get
\begin{equation}d_0(x_0,x_{l'})\leq l'+2  \end{equation}
\begin{lemma}For two arbitrary nodes
\begin{equation}x_0\in C_0,x_{l'}\in
  C_{l'}\;\text{with}\;d_1(C_0,C_{l'})\leq l   \end{equation}
we have
\begin{equation}d_0(x_0,x_{l'})\leq l'+2  \end{equation}
and hence
\begin{equation}|U_l^{cl}(C_0)|\subset U_{l+2}(x_0)   \end{equation}
with $|U_l^{cl}(C_0)|$ the set of $0$-nodes, lying in $U_l^{cl}(C_0)$
(the latter set now understood as the set of its 0-nodes). This
implies
\begin{equation}\#(|U_l^{cl}(C_0)|)\leq\#(U_{l+2}(x_0))    \end{equation}
\end{lemma}

From observation 4.2 of \cite{4} we know that each node, $x_i$, can
lie in at most $2^{v_i}$ different cliques, with $v_i\leq v$. This
yields the crude, but apriori estimate
\begin{equation}\#(U_l^{cl}(C_0))\leq\#(U_{l+2}(x_0))\cdot 2^v   \end{equation}
which is the desired upper bound on the number of cliques, lying in
$U_l^{cl}(C_0)$. We conclude that, for an infinite graph with $v_i\leq
v<\infty$, we have for the dimension of its clique graph:
\begin{equation}\overline{D}_{cl}\leq D   \end{equation}
since 
\begin{equation}\ln(\#(U_l^{cl}(C_0)))/\ln(l)\leq \ln(\#(U_{l+2}(n_0)))/\ln(l)+v\cdot\ln(2)/\ln(l)     \end{equation}
For $l\to\infty$ we get the above result.

We want to prove a corresponding lower bound. Take an arbitrary node,
$x_{l'}$, in $U_l(x_0)$. By definition, there exists a node-
(edge-)sequence
\begin{equation} \label{ordinary}x_0-x_1-\cdots-x_{l'}\;\text{with}\;l'\leq l    \end{equation}
On the other side, there exists a sequence of cliques, $C_i$, with
each consecutive pair of nodes, $(x_{i-1},x_i)\in C_i$. These cliques
do exist because, starting from the connected pair,$(x_{i-1},x_i)$, we
get such a clique by extending this germ in one of (possibly) several
ways to a clique (cf. section 4 of \cite{4}). We can conclude that for
each node, $x_{l'}\in U_l(x_0)$, and $x_0\in C_0$, we have
\begin{equation}x_{l'}\in|U^{cl}_{l+1}(C_0)|    \end{equation}  
(note that the clique, containing both $x_0$ and $x_1$ may be different
from the start clique, $C_0$!).

We then have
\begin{equation}U_l(x_0)\subset |U^{cl}_{l+1}(C_0)|\quad\text{and}
\quad\#(U_l(x_0))\leq\#(|U^{cl}_{l+1}(C_0)|)   \end{equation}
With $v_i\leq v$ for all $x_i$, the maximal order of a clique is
bounded from above by $(v+1)$. This implies
\begin{equation}\#(|U^{cl}_{l+1}(C_0)|)\leq (v+1)\cdot\#(U_{l+1}^{cl}(C_0))    \end{equation}
and
\begin{equation}\#(U_{l+1}^{cl}(C_0))\geq \#(U_l(x_0))/(v+1)  \end{equation}
We hence get
\begin{equation}\ln(\#(U_{l+1}^{cl}(C_0)))/\ln(l+1)\geq \ln(\#(U_l(x_0)))/\ln(l+1)-\ln(v+1)/\ln(l+1)    \end{equation}
With 
\begin{equation}\ln(l+1)=\ln(l\cdot(1+l^{-1}))=\ln(l)+\ln(1+l^{-1})    \end{equation}
and $l\to\infty$, we see that
\begin{equation}\underline{D}_{cl}\geq D   \end{equation}
and get the important theorem:
\begin{satz}Assuming that $G$ has dimension $D$ and globally bounded
  node degree, $v_i\leq v<\infty$, we have that $D_{cl}$ also exists
  and it holds
\begin{equation}D_{cl}=D   \end{equation}
Note that this result does hold for the ordinary clique graph, viz.
arbitrary overlap, viz., no purification. In other words, under these
assumptions, the renormalisation steps do not change the graph
dimension.
\end{satz}
This result is reminiscent of a similar observation in statistical
mechanics where the non-coarse-grained Gibbsian entropy happens to be
a \tit{constant of motion}. The same happens here. In the ordinary
clique graph each original bond occurs in at least one clique, i.e.
there is no real (or, more precisely, not enough) coarse graining.
\section{Critical Network States}
In subsection \ref{ana} we derived formulas for the size of the
so-called local group of a clique in a random graph, that is the set
of cliques with which a given clique has a (sufficient) common
overlap.  If one is in the parameter regime in which the cliques are
still densely and complicately entangled (typically the first
renormalisation steps) and compares the number of bonds in the
\tit{purified} clique graph, that is, bonds being defined by a
sufficient! overlap, with the number of bonds in the corresponding
(unpurified) clique graph, the latter number exceeds the former one by
many orders. Put differently, in this situation the number of
\tit{marginal} overlaps of cliques is much bigger. All these marginal
overlaps are deleted in the purification or renormalisation process.

The last theorem in the preceding section shows that we will not get a
dimensional reduction without sufficient purification. If we go
through the proof we see that the first part does hold unaltered for
the purified clique graph. In the second part, however, we used an
argument which does only hold for ordinary clique graphs (see the
remarks following formula (\ref{ordinary}). The existence of the row
of overlapping cliques, employed there, can only be guaranteed if the
degree of overlaps are left arbitrary. We hencecan infer:
\begin{koro}For the purified clique graph, with overlaps exceeding a
  certain fixed number, $l_0$, we can only prove
\begin{equation}D_{cl}\leq D\end{equation}
\end{koro}
Having for example the picture in mind, frequently invoked by Wheeler
and others, of a space-time foam, with a concept of dimension
depending on the scale of resolution (see e.g. Box 44.4 on p.1205 in
\cite{Wheeler}), we infer from our above observations that this may
turn out to be both an interesting and not entirely trivial topic. We
have to analyze under what specific conditions the dimension can
actually shrink under coarse-graining, so that we may start from a
very erratic network on, say, the Planck scale, and arrive in the end
at a smooth macroscopic space-time having perhaps an integer dimension
of, preferably, value 4 or so.

We remarked already in the introduction that geometric change or
geometric phase transitions are supposed to be related to some sort of
\tit{critical state} of the network. Our previous observations about
the possibility of dimensional change under coarse graining together
with an interesting observation already made in \cite{1}, lemma 4.10,
allows us to \tit{almost rigorously} prove what kind of criticality is in
fact necessary to achieve this goal.

We showed there that it is not so easy to modify the dimension of a
graph by \tit{local} alterations.   
\begin{propo}Additional insertions of bonds between arbitrarily many nodes,
  $y,z$, having original graph distance, $d(y,z)\leq k\;,\;k\in\N$
  arbitrary but fixed, do not change $\underline{D}(x)$ or
  $\overline{D}(x)$.
\end{propo}
From this we learn the following. Phase transitions in graphs,
changing the dimension, have to be intrinsically \tit{non-local}. That
is, they necessarily involve nodes, having an arbitrarily large
distance in the original graph. We think, this is a crucial
observation from the physical point of view. On the one side, it shows
that systems have to be \tit{critical} in a peculiar way, that is,
having a lot of distant correlations or, rather, correlations on all
scales (cf. also Smolins's discussion in e.g. \cite{Smo1} and
elsewhere). On the other side, it fits exactly with our working
philosphy that quantum theory is a \tit{residual} and \tit{coarse
  grained} effect of such largely hidden long range correlations
(\cite{Quantum}).

If we apply these findings to our renormalisation steps, that is,
passing from a graph to its associated (purified) clique graph, this
implies the following. We saw that assuming a network or graph, $G$,
having a dimension, $D$, the unpurified clique graph still has 
\begin{equation}D_{cl}=D\end{equation}
On the other hand, denoting the purified clique graph by
$\hat{G}_{cl}$, we have the estimate
\begin{equation}\hat{D}_{cl}\leq D_{cl}=D    \end{equation}

The transition from $G_{cl}$ to $\hat{G}_{cl}$ consists of the
deletion of marginal overlaps among cliques (with the necessary
criteria provided by the physical context). That is, $\hat{G}_{cl}$
lives on the same node set (the set of cliques) but has fewer
(meta)bonds. The above proposition shows that this does \tit{not}
automatically guarantee that we really have 
\begin{equation}\hat{D}_{cl}< D_{cl}   \end{equation}
Quite to the contrary, we learned that this can only be achieved if
the bond deletions happen in a very specific way.

On $G_{cl}$ we have, as on any graph, a natural distance or
neighborhood structure, given by the canonical graph metric,
$d_{cl}(C_i,C_j)$. Note that the above proposition holds as well for
bond deletions instead of insertions. We thus infer that bond
deletions in $G_{cl}$ between cliques which are not very far apart
 in the final purified graph $\hat{G}_{cl}$ cannot alter the final
 dimension of $\hat{G}_{cl}$. More precisely, only bond deletions
 between cliques having distances in $\hat{G}_{cl}$ which approach
 infinity in a specific way, can have an effect.
\begin{conclusion}We conclude that only the bond deletions between
  very distant cliques (with respect to $\hat{G}_{cl}$), with this
  distance being unbounded, can decrease the dimension of $\hat{G}_{cl}$
as compared to $G_{cl}$. More precisely, there has to be a
substantial bond deletion on all scales up to infinity.
\end{conclusion}
The above observation reminds one of the \tit{scale invariance} of
\tit{critical systems} in other contexts. We exemplify this by a
simple but instructive example.

This (inhomogeneous; it slightly depends on the reference point $(0,0)$)
construction has already been given in section 5.2.5 of \cite{1}.  One
takes the lattice, $\Z_2$, inscribes in it, starting from the point
$(0,0)$, two non-intersecting outwardly spiraling edge sequences:
\begin{equation}(0,0)\to(1,0)\to(1,1)\to(0,1)\to(-1,+1)\to(-2,+1)\to(-2,0)\to\cdots\end{equation}
and
\begin{equation}(0,0)\to(-1,0)\to(-1,-1)\to(0,-1)\to(+1,-1)\to(+2,-1)\to(+2,0)\to\cdots
\end{equation}
\begin{figure}[h]
\centerline{\epsfig{file=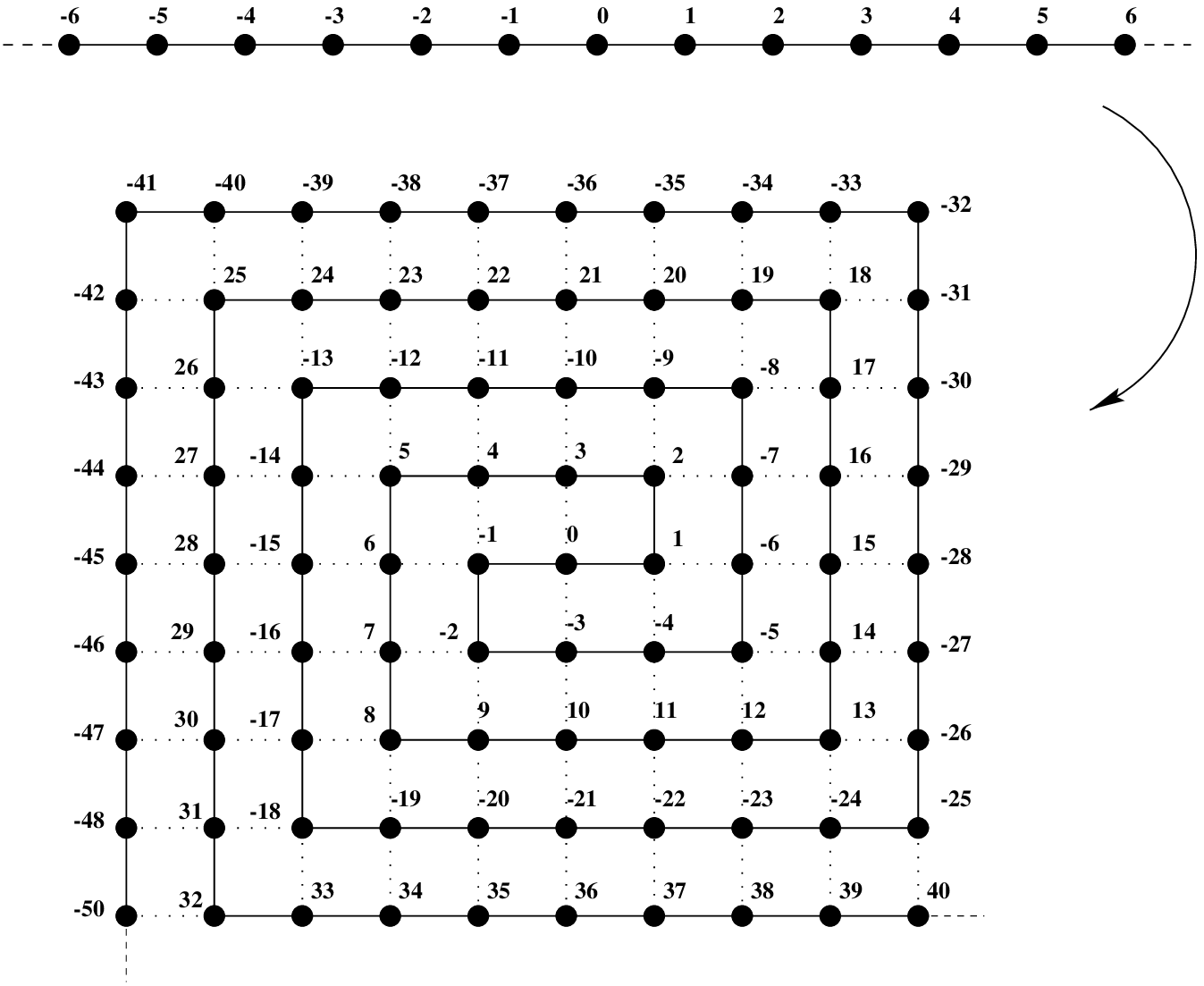,width=8cm,height=8cm,angle=0}}
\caption{}
\end{figure}
We consider this inscribed graph as a representation of the
one-dimensional lattice, $\Z_1$, with the node labelling running from $-\infty$
to $+\infty$.
\begin{equation}0\to1\to2\to3\to4\to5\to\cdots \end{equation}
and
\begin{equation}0\to-1\to-2\to-3\to-4\to-5\to\cdots    \end{equation}
\begin{bem} The embedded graph, being isomorphic to $\Z_1$, is in fact a
  {\em spanning tree} relative to the ambient graph, $\Z_2$.
\end{bem}

One can now see that the extra bonds, occurring in $\Z_2$, not
belonging to the representation of $\Z_1$, connect nodes of a larger
and larger distance with respect to the labelling of $\Z_1$. We have
for example bonds in $\Z_2$ between pairs of nodes with the $\Z_1$-labels,
\begin{equation}0,3\,;\,3,-10\,;\,-10,21\,;\,21,-36\,\ldots    \end{equation}
and correspondingly for other sequences of nodes. That is, the
embedded graph is one-dimensional, lying in a two-dimensional graph,
while the node sets are identical. The preceding discussion and the
figure illustrate and confirm what we have said above about the type
of necessary criticality and long-range correlations.

To employ this example for our renormalisation group approach, we can
replace the original nodes (with the $\Z_1$-labelling) by certain
cliques of arbitrary order, choose the overlaps appropriately, so that
the above representation of $\Z_1$ becomes the \tit{purified} clique
graph of the total graph. We arrive at a coarse-grained graph of
dimension one, starting from an unpurified graph of dimension two or a
larger dimension.

\end{document}